\documentclass[lineno]{JFM-FLM_Au}

\newcommand{\AN}[1]{\textcolor{black}{#1}}
\newcommand{\R}[1]{\textcolor{black}{#1}}
\newcommand{\RR}[1]{\textcolor{black}{#1}}
\usepackage{changes}
\usepackage{comment}

\lefttitle{Amir Nazemi and Hongyi Xiao}
\righttitle{Journal of Fluid Mechanics}

\title{Reciprocal swimming in granular media: the role of jamming and swimmer inertia}

\author{Amir Nazemi\aff{1} \and Hongyi Xiao\aff{1}}

\affiliation{\aff{1}Department of Mechanical Engineering, University of Michigan, Ann Arbor, USA}

\corresau{Hongyi Xiao, \email{hongyix@umich.edu}}

\begin{document}
\nolinenumbers
\maketitle

\begin{abstract}
We use particle simulations to reveal two distinct propulsion mechanisms for a scallop-like swimmer to propel itself in granular media by reciprocally flapping its wings.
\R{Based on the discrete element method, we examine the structure, kinematics, and contact forces of particles near the swimmer to quantify how jamming manifests as stagnant zones near the swimmer in a frictional granular medium, which are less intense during the opening stroke than the closing.} 
This broken symmetry is quantified by the difference in the number of strong particle contact forces formed during opening and closing, which shows a linear relation with the swimmer’s net displacement across various swimmer and medium configurations, all favoring the opening stroke.
We identify a secondary propulsion mechanism in a dynamic regime with significant swimmer inertia, as the flapping period approaches the coasting time for a moving swimmer to come to rest under the medium resistance. In this case, the swimmer’s net displacement is correlated to the ratio between these two time scales, and the swimming direction favors the closing stroke due to the smaller medium resistance as the swimmer coasts with closed wings.

\end{abstract}


\section{Introduction}
\label{sec:intro}


Locomotion in granular media is challenging because granular particles can both flow like a complex fluid and jam into a solid~\citep{jaeger1996granular,de1999granular}. To navigate granular media, many swimmers adopt intricate gaits~\citep{hosoi2015beneath,aguilar2016review,kamrin2024advances},
such as undulation \citep{maladen2009undulatory, maladen2011mechanical, ding2012mechanics, peng2016characteristics, schiebel2020mitigating, li2024enhancing} and peristaltic body expansion~\citep{winter2014razor, dorgan2015biomechanics, isaka2019development, zhang2023bioinspired, dorgan2023fundamentals}.
On the other hand, gaits as simple as flapping an appendage also prove effective for granular locomotion of animals and bio-inspired robots~\citep{holm1973daily,zhang2019motion,chopra2020granular,li2021compliant,darbois2021propulsion,xiao2024locomotion}.
By periodically oscillating rigid appendages such as rods and plates, swimmers can consistently propel themselves in granular media~\citep{darbois2021propulsion,xiao2024locomotion}. 
This is likely not predictable by simplified models such as certain resistive force theories~\citep{li2013terradynamics}, which assume time-reversal symmetry and often neglect the evolving state of the granular medium, e.g., due to the jamming effects~\citep{bi2011jamming}. 
This highlights the need to better understand locomotion mechanisms that utilize the complexity of the medium.

Swimming with reciprocally flapping wings recalls the scallop theorem~\citep{purcell2014life}, which states that reciprocal motion, consisting of time-reversible body deformation sequences, cannot produce net displacement in Newtonian fluids at low Reynolds numbers.
\R{For Newtonian fluids, this topic has been investigated extensively for biological locomotion~\citep{shapere1987self,shapere1989geometry,koiller1996problems} and engineering applications~\citep{wang2018design, wang2020development, robertson2019roboscallop}, resulting in various strategies to break the time-reversal symmetry~\citep{lighthill1976flagellar, brennen1977fluid, becker2003self, lauga2008no,lauga2011life}. }
\R{In granular media, however, time-reversal symmetry generally does not hold, appearing only in specialized configurations~\citep{otsuki2021shear,zhao2022ultrastable}.} \R{Consequently, the scallop theorem is inapplicable, as the complexity of granular locomotion violates two of its preconditions.} 

First, granular media have strongly non-Newtonian rheology~\citep{gdr2004dense,jop2006constitutive,kamrin2012nonlocal}, \R{and granular locomotion often occurs in a regime where the medium experiences elasto-plastic deformation that is rate-independent~\citep{hosoi2015beneath}}. Near an intruding object, e.g., a swimmer, jamming effects in the medium manifest, which often result in the formation of stagnant zones with jammed particles that move along with the intruder without rearranging~\citep{aguilar2016robophysical, kang2018archimedes, feng2019support, agarwal2019modeling, harrington2020stagnant, pravin2021effect, agarwal2021surprising, agarwal2023mechanistic, yin2025extended}. 
\R{As a result, these jammed particles effectively become an internal part of the intruder, enlarging its body size.}
\R{The stagnant zones may require a finite amount of intruder displacement to develop~\citep{feng2019support,harrington2020stagnant}, inducing history-dependence in resistive forces.}
More complexities may arise due to overlapping length scales of the particle size, stagnant zone size and intruder size~\citep{pravin2021effect, kozlowski2019dynamics}. 
\R{These complexities require granular locomotion models to incorporate more physics such as history-dependent medium response and elasto-plasticity~\citep{Yilmaz2026ElasticRFT}. }

Second, the swimmer's inertia can play a significant role in granular swimming, even when the medium exhibits rate-independent deformation.
For a typical swimmer like a bacterium in a liquid at low Reynolds numbers, both the swimmer's and the fluid's inertia are negligible due to their similar densities.
For granular locomotion, however, a swimmer denser than the surrounding particles can possibly be sustained without sinking, especially given that the medium tends to exert lift forces on laterally moving intruders~\citep{ding2011drag,chopra2020granular}. 
For a swimmer with non-negligible inertia, reciprocal strokes could potentially break the time-reversal symmetry and enable locomotion~\citep{gonzalez2009reciprocal}.
While this mechanism is under-explored in granular media, it has been demonstrated using a dense interfacial swimmer whose weight is supported by surface tension \citep{hubert2021scallop}. In this case, the swimmer's coasting time, the characteristic duration over which fluid drag halts its motion after a stroke, becomes comparable to the driving period. When the respective coasting times for the two reciprocal strokes differ, net locomotion occurs even if the surrounding fluid remains in the Stokes regime~\citep{gonzalez2009reciprocal}.
This type of inertial effect remains to be investigated in the context of granular locomotion.

In this study, we seek to reveal the respective roles of jamming and swimmer inertia in reciprocal granular swimming using discrete element method (DEM) simulations,  
building upon our recent experiments with a scallop-like robot that swims in granular media by reciprocally flapping its wings~\citep{xiao2024locomotion}.
\R{In quasi-static swimming, we use microscopic structural and force transmission analyses to show how the history-dependent contact network reorganization in the jammed granular medium breaks the symmetry between the opening and closing strokes, which is reflected by the formation of asymmetric stagnant zones near the swimmer under reciprocal actuation.} 
\RR{The coarse-grained particle velocity field further shows the asymmetry in the swimmer-induced particle kinematics.}
\R{The implications of stagnant zone formation on swimming performance are further demonstrated through simulations with varied swimmer geometries and particle friction coefficients. These results quantitatively establish why net propulsion is biased towards the opening stroke as observed in experiments~\citep{xiao2024locomotion}.}
The swimmer's inertia becomes significant in dynamic swimming, e.g., with faster flapping frequencies, where the swimmer's coasting time approaches the wings' flapping period. The swimmer experiences a longer coasting time with closed wings, resulting in a reversal in propulsion direction, which favors the closing stroke. 
For both mechanisms, we establish physics-based order parameters that strongly correlate with the respective net locomotion displacements.

\section{Simulation Method}\label{sec:methods}

\begin{figure}[t]
    \centering
    \includegraphics[width=\linewidth]{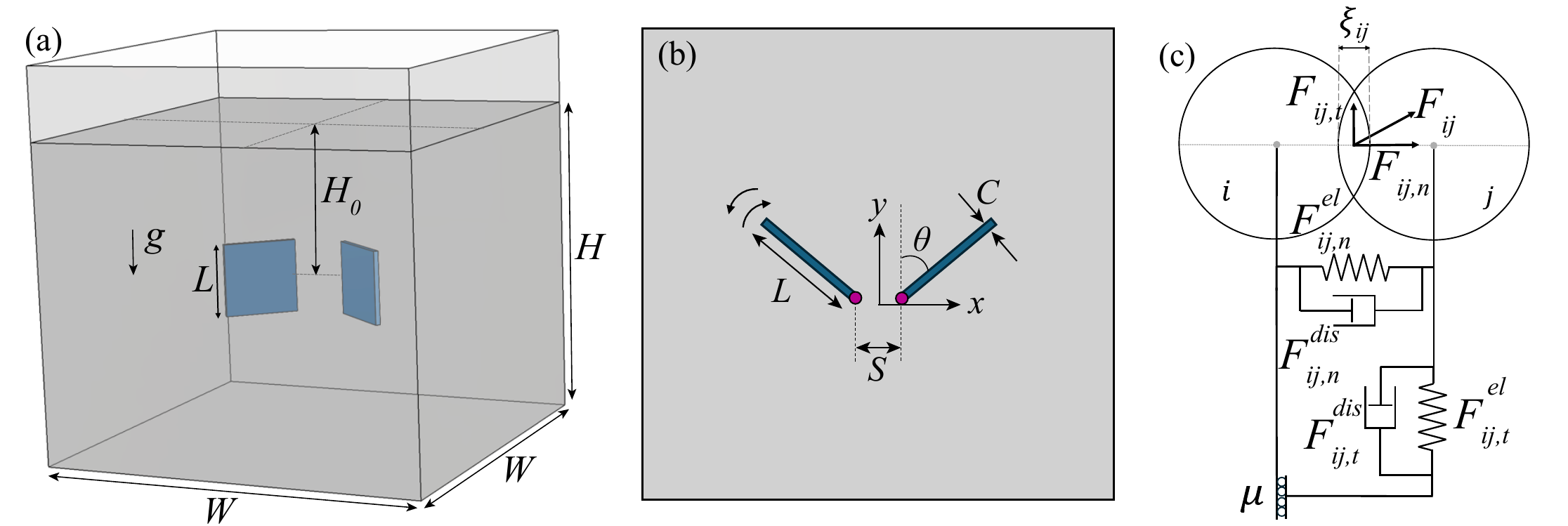}
    \caption{Schematics of the simulation setup and particle contact models. (a) 3D view of the setup. (b) Top-down view of the setup. (c) Normal and tangential contact forces between particles $i$ and $j$.}
    \vspace{-5 mm}
    \label{fig:setup}
\end{figure}

We simulate a scallop-like swimmer \R{submerged at depth $H_0$ measured from the free surface} of a granular reservoir, see Fig.~\ref{fig:setup}a. 
The swimmer has a pair of square-shaped wings with length $L$ and thickness $C$. The wings rotate cyclically around their respective hubs (purple dots in Fig.~\ref{fig:setup}b), which are set at a gap width $S$ to each other. 
A coordinate system is established, with the origin set at the midpoint between the two hubs, the $x$ direction pointing laterally, the $y$ direction aligned with the swimmer translation, and the $z$ direction against the gravitational acceleration $g$.
We define the wing angle, $\theta$, as the angle between the $y$-axis and a vector in the $x$-$y$ plane tangent to a wing's surface.
In each simulation, the swimmer reciprocally opens and closes its wings with a period $T$, and the range of $\theta$ is bounded by the maximum and minimum angles, $\theta_{\mathrm{o}}$ and $\theta_{\mathrm{c}}$, respectively.
The wings rotate at a constant angular speed, $\omega=2(\theta_{\mathrm{o}}-\theta_{\mathrm{c}})/T$. In a subset of simulations, we also consider the swimmer to have inert components, as discussed later.

We model the interaction between a pair of contacting particles, $i$ and $j$, with a force, \(\boldsymbol{F}_{ij}=\boldsymbol{F}_{ij,n}+\boldsymbol{F}_{ij,t}\), as depicted in Fig.~\ref{fig:setup}c.
The normal component is $\boldsymbol{F}_{ij,n} = (F_{ij,n}^{\text{el}}+F_{ij,n}^{\text{dis}})\boldsymbol{n}_{ij}$ and the tangential component is $\boldsymbol{F}_{ij,t} = F_{ij,t}\boldsymbol{t}_{ij}$, where $\boldsymbol{n}_{ij}$ is the unit vector connecting particle centroids and $\boldsymbol{t}_{ij}$ is the tangential unit vector, set by the direction of the relative tangential velocity between $i$ and $j$~\citep{thornton2015granular,thornton2023recent}.
For the scalar values of the normal force, the Hertz model~\citep{hertz1882z} was used to calculate the elastic component, and a viscous model~\citep{muller2011collision} was used for dissipation, which are

\vspace{-2 mm}
\begin{equation}
    F_{ij,n}^{\text{el}}=\frac{2}{3}\frac{E}{1-\nu^2}\sqrt{R_{\text{eff}}}\xi_{ij}^{3/2}
    \label{eq:4},
\end{equation}
\begin{equation}
    F_{ij,n}^{\text{dis}}=\gamma_n\dot{\xi}_{ij}=\frac{E}{1-\nu^2}\sqrt{R_{\text{eff}}}A_{\gamma}\sqrt{\xi_{ij}}\dot{\xi}_{ij}
    \label{eq:5},
\end{equation}

\noindent where $E$ is the Young's modulus, $\nu$ is the Poisson's ratio, \(R_{\text{eff}}=\frac{R_iR_j}{R_i+R_j}\) is the effective particle radius, \(\xi_{ij}=R_i+R_j-|\boldsymbol{r}_i-\boldsymbol{r}_j|\) is the overlap in the normal direction, $\dot{\xi}_{ij}$ is the relative normal velocity, $R_i$ is the particle radius, $\boldsymbol{r}_i$ is the particle position, and $A_{\gamma}$ is a dissipation constant that can be set according to \cite{muller2011collision}\R{, which is set corresponding to a restitution coefficient of 0.7 at a characteristic relative velocity of $4$\,mm/s, reflecting the swimmer's wing velocity.} 
The tangential interaction combines a viscoelastic force with Coulomb friction \citep{mindlin1949compliance,roy2024combined,roy2024structural}, which is

\vspace{-2 mm}
\begin{eqnarray}
    F_{ij,t} =\left\{
    \begin{array}{l l}
        \mu F_{ij,n}                     &   \quad \mu F_{ij,n} \le F_{ij,t}^{\mathrm{el}}+F_{ij,t}^{\mathrm{dis}}\\
        F_{ij,t}^{\mathrm{el}} + F_{ij,t}^{\mathrm{dis}}   &   \quad \mu F_{ij,n}>F_{ij,t}^{\mathrm{el}}+F_{ij,t}^{\mathrm{dis}}
    \end{array}\right.
    \label{eq:7},
\end{eqnarray}
\begin{eqnarray}
    F_{ij,t}^{\mathrm{el}}(\delta_{ij})
    =\frac{4}{3}\frac{E}{1-\nu^2}\sqrt{R_{\text{eff}}}\sqrt{\xi_{ij}}\delta_{ij}
\label{eq:8},
\end{eqnarray}
\begin{eqnarray}
    F_{ij,t}^{\mathrm{dis}}=\gamma_t\dot{\delta}_{ij}, \quad \quad \gamma_t=\gamma_n
    \label{eq:9},
\end{eqnarray}

\noindent where $\mu$ is the friction coefficient, $\delta_{ij}$ is the accumulated relative tangential displacement between particles $i$ and $j$, and $\dot{\delta}_{ij}$ is the relative tangential velocity.
We adopted the DEM package \textit{MercuryDPM} \citep{WEINHART2020107129, thornton2023recent} and simulated approximately 510,000 particles having a density of $\rho=1050$\,kg/m$^3$, and a diameter of $d=1.0\pm 0.12$\,mm. The lateral dimensions of the reservoir are then $W=80d$.
We used a reduced Young's modulus so that a large time step can be used \citep{brilliantov1996model}. \R{Using a Young's modulus that is 10$\times$ larger would slightly reduce the average number of contacts per particle, while having little influence on the observed locomotion.} We selected a common value of $\mu=0.4$ for friction as a base value, but we will test the influence of $\mu$ later on.
The bottom and side walls of the reservoir are frictional boundaries as in our previous experiment \citep{xiao2024locomotion}. 

\begin{table}
    \centering
    \def~{\hphantom{0}}
        \begin{tabular}{l l c}
            Parameter & Symbol (unit) & Value \\[3pt]
            mean particle diameter & $d$ (mm) & 1.0 \\
            wing length & $L$ (mm) & 15.0 \\
            wing thickness & $C$ (mm) & 1.0 \\
            wing gap width & $S$ (mm) & \{0.0, 2.75, 5.5, \textbf{11.0}, 22.0\} \\
            container width & $W$ (mm) & 80.0 \\
            free surface height & $H$ (mm) & 65.0 \\
            swimmer depth & \R{$H_0$} (mm) & 32.5 \\
            closed angle & $\theta_{\mathrm{c}}$ (deg) & 20 \\
            open angle & $\theta_{\mathrm{o}}$ (deg) & 80 \\
            swimming period & $T$ (s) & \{0.05, 0.1, 0.2, 0.5, 1.0, \textbf{2.0}\} \\
            Young's modulus & $E$ (MPa) & 1.0 \\
            Poisson's ratio & $\nu$ & 0.34 \\
            friction coefficient & $\mu$ & \{0.0, \textbf{0.4}\} \\
            dissipation constant & $A_{\gamma}$ (s) & $1.63\times10^{-5}$ \\
            particle density & $\rho$ (kg$/$m$^3$) & 1050 \\
            swimmer mass & $m_{s}$ (g) & \{0.012, \textbf{0.60}, 1.19, 2.38, 9.53, 29.78, 1370.00\} \\
            time step size & $\Delta t$ (s) & $1\times10^{-5}$ \\
        \end{tabular}
    \vspace{-2mm}
    \caption{\label{tab:1} Parameters used in the DEM simulation. Bold values are used in the base case.}
\end{table}


As for the swimmer, its wings interact with particles with the standard particle-boundary interaction, having identical properties as in the particle-particle interaction. We impose the rotation on the wings and integrate the swimmer's translational motion using the Verlet scheme \citep{verlet1967computer}. 
In our experimental setting \citep{xiao2024locomotion}, the swimmer's position was fixed, and the reservoir was lubricated so that it could move in the $y$ direction due to the swimming force. Accordingly, the simulated swimmer has only one translational degree of freedom in $y$, while its mass ($m_s$) is set to be concentrated at the hubs. 
A parametric study was performed with systematically varied parameters as listed in table \ref{tab:1}.

\R{In all simulations, 
particles were inserted above the swimmer, and they settled down by gravity. A particle deletion boundary was further introduced above the reservoir, which slowly descends on the reservoir's free surface and deletes all particles it contacts. This helps set a flat free surface at the desired height $H=65d$. After an additional period for the system to equilibrate, we simulate five reciprocal swimming cycles.}

\section{Results}
\subsection{Swimmer and Particle Kinematics}\label{sec:val}

\begin{figure}[t]
    \centering
    \includegraphics[width=\linewidth]{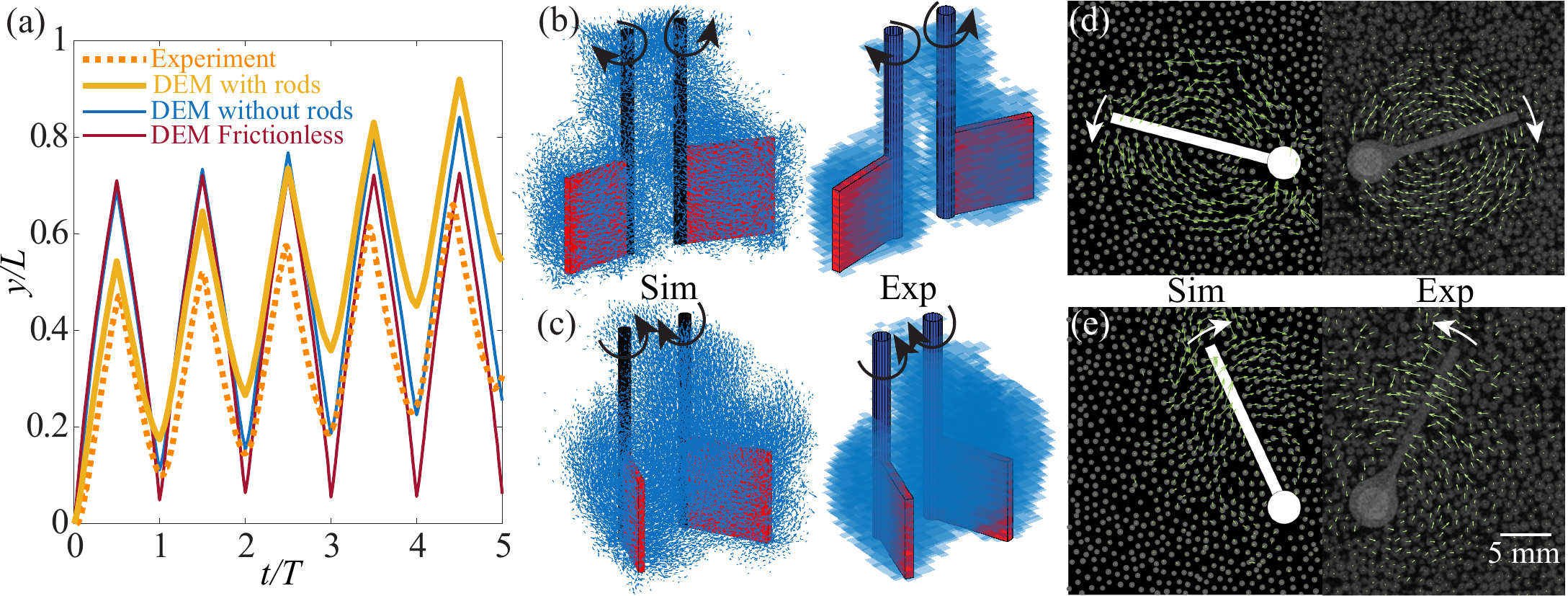}
    \vspace{-1 mm}
    \caption{Swimmer and particle kinematics from the experiment \citep{xiao2024locomotion} and from DEM simulations, see configurations in table \ref{tab:1}. (a) Swimmer displacement vs time in the experiment (dotted, orange), the DEM result with simulated mounting rods (yellow), the simplified simulation \textit{base case} (blue), and a frictionless simulation (red). 
    (b) and (c) Highlighting particles with a displacement larger than five times the average displacement in the DEM results (left panels) and the experimental X-ray CT results (right panels), at $\theta=75^\circ$ during opening and $\theta=25^\circ$ during closing, respectively. (d) and (e) Comparison of the particle displacement field at the mid-plane of the swimmer between DEM results (left panels) and X-ray CT results (right panels), at $\theta=75^\circ$ during opening and $\theta=25^\circ$ during closing, respectively.}
    \label{fig:loc}
    \vspace{-3 mm}
\end{figure}

We first compare the simulated swimmer locomotion and particle kinematics with results from experiments that have identical setups in the swimmer configuration and reservoir size, as described in our previous work \citep{xiao2024locomotion}. \R{In the experiment, we observed a persistent propulsion over many cycles.}
As the experimental swimmer was mounted on two vertical rods, we included the rods in one DEM simulation for validation.
Figure~\ref{fig:loc}a shows the normalized displacement of the swimmer, $y/L$, over normalized time, $t/T$. The experimental measurement (dotted curve) and the corresponding simulation result (thick yellow curve) show qualitative agreement, with consistent net locomotion in the $+y$ direction and similar displacement amplitudes per stroke. The simulated net displacement per cycle is larger than that in the experiment. This is likely due to a higher initial packing fraction, $\phi_0=0.60$, in the simulation, compared to the experiment ($\phi_0\approx0.58$), as the experimental particle system was air-fluidized before each experiment~\citep{xiao2024locomotion}, while particles settle under gravity in DEM.
We also plot the result of a simulation without the rods (blue curve), which shows the same $+y$ locomotion direction, a bigger displacement amplitude per stroke, and less net locomotion in each cycle.
For simplicity, we will later focus on this simulation in the following sections and refer to it as the \textit{base case}. 


\begin{figure}[t]
    \centering
    \includegraphics[width=\linewidth]{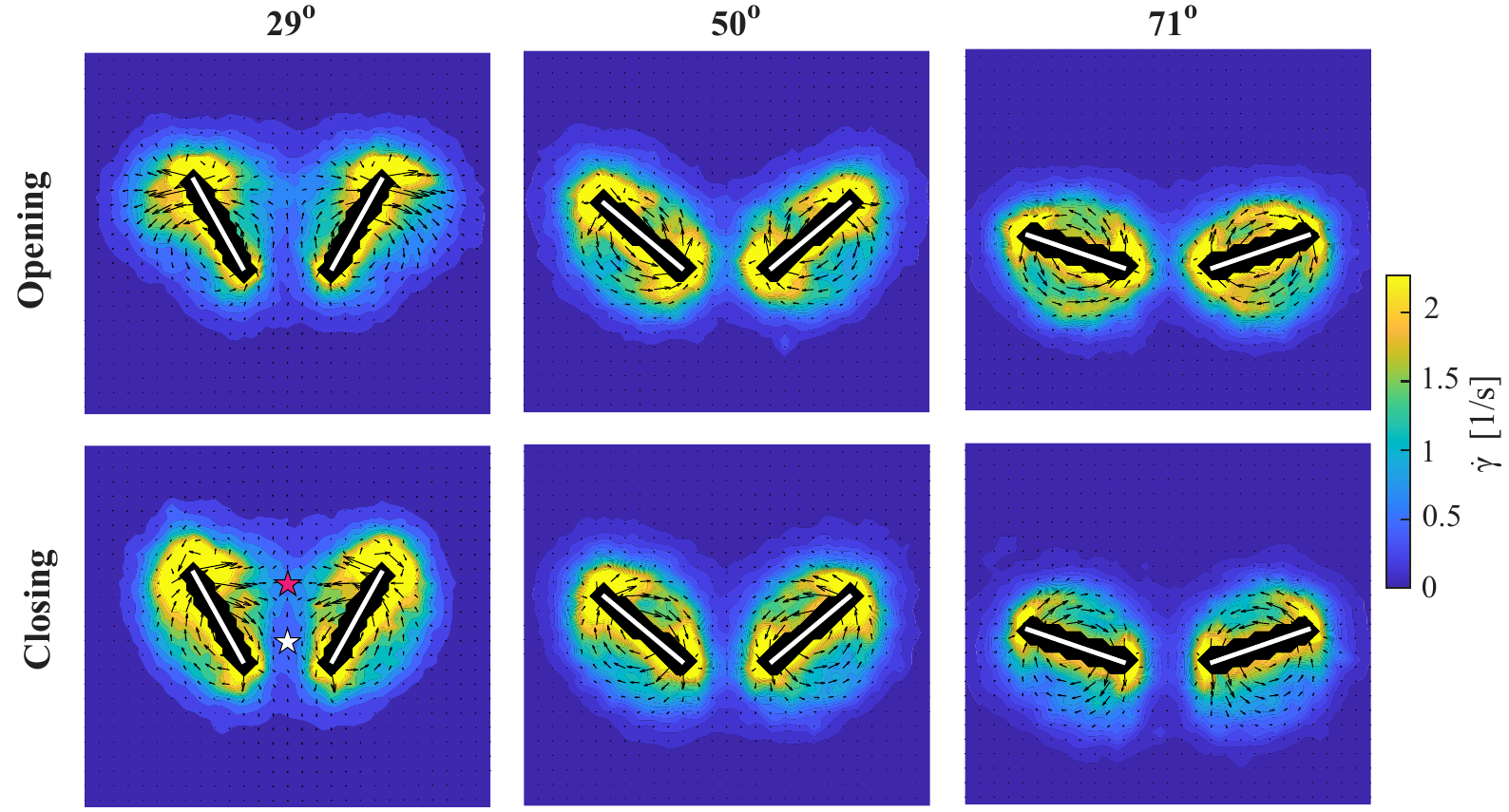}
    \vspace{-5 mm}
    \caption{\RR{The coarse-grained flow field in the swimmer's mid-plane, with velocity vectors and shear rate $\dot{\gamma}$ visualized for the frictional base case. The three columns correspond to $\theta = 29^\circ, \, 50^\circ, 71^\circ$, respectively, while the two rows correspond to the opening and closing stroke, respectively. The black region covers the bins that overlap with the wings (white). The purple star indicates a highly crowded region and the white star indicates a depleted region. 
    }}
    \label{fig:flow}
    \vspace{-4 mm}
\end{figure}

To continue our validation, we visualize particle kinematics using DEM and a quasi-static experiment imaged by X-ray computed tomography (CT)~\citep{xiao2024locomotion}, capturing one full 3D scan of the system after an incremental wing rotation.
In Fig. \ref{fig:loc}b-e, we first highlight regions with significant particle displacement (panels b and c), and then plot the in-plane ($x$-$y$) displacement field in the mid-plane of the swimmer (panels d and e), both showing good agreement between the simulation and experiment. 
\RR{Besides the individual particle displacement, we obtained the coarse-grained velocity field of the particles around the swimmer using a binning method, in which the average velocity vector in each bin is weighted by the exact partial volumes of the spheres overlapping with the bin~\citep{strobl2016exact}. This allows us to use smaller bins for better spatial resolution, and we further average the binning results over five swimming cycles. Results are displayed in Fig.~\ref{fig:flow} for three example angles during the opening and closing stroke, showing the vector field of the $x,y$ component of the particle velocity, along with underlying contours, which represent the strain rate, $\dot{\gamma}$, calculated as the second invariant of the deviatoric part of the 3D strain rate tensor. }

\RR{At intermediate and large opening angles, $\theta=75^{\circ}$ in Fig.~\ref{fig:loc}d and $\theta=50^{\circ}, 71^{\circ}$ in Fig.~\ref{fig:flow}, each wing induces a rather localized flow in its immediate surroundings, with the radius of the perturbed region comparable to the wing's length, and it does not overlap with the other wing's perturbed region. More specifically, the flow around each wing organizes into a vortex, in which particles follow the wing rotation as if they are part of the wing. This is similar to the jammed stagnant zones observed in previous studies of penetrating plates~\citep{kang2018archimedes,feng2019support,harrington2020stagnant}, which are cone- or triangular-shaped zones appearing in front of the leading edge. Here, we also see that the strain rate is smaller towards the center of the wings, reflecting the cone shape. However, we note that for intruders that are not much larger than the particle diameter, deviatoric deformation and particle rearrangement can still occur~\citep{harrington2020stagnant}, and particles do not move as a perfectly rigid cluster. While this adds difficulty in detecting the stagnant zones, we will use force-based methods to better characterize them in Sec.~\ref{sec:jam}. With the co-moving stagnant zones, the swimmer-induced flow field is reminiscent of that produced by a counter-rotating vortex pair in a Newtonian liquid, such as the flow generated by a pair of counter-rotating cylinders~\citep{van2015self}. While classical fluid dynamics establishes that the propulsion speed of such a vortex pair is linear in the vortex strength~\citep{batchelor2000introduction}, the analogous behavior in a granular medium remains unexplored.}

\RR{At smaller angles, $\theta=25^{\circ}$ in Fig.~\ref{fig:loc}e and $\theta=29^{\circ}$ in Fig.~\ref{fig:flow}, the vortex-like flow field no longer exists, especially during closing, and the two zones perturbed by the wings
start to overlap. Near the center of the swimmer, we see the in-plane velocity starts to converge, which in turn triggers out-of-plane particle motion, see Fig.~\ref{fig:loc}c. 
In Appendix~\ref{app:3dvf}, we further show the corresponding velocity fields in 3D, confirming a strong upward flow during closing and a downward flow during opening at $\theta=29^\circ$. At larger $\theta$, the particle motion is mostly in-plane. 
The flow field at smaller $\theta$ is also more extensive than that at larger $\theta$, which indicates that localized particle rearrangements become more difficult, signaling intensified jamming effects~\citep{xiao2024locomotion}. We next examine the role of jamming in greater detail using force transmission and structural information.
}

\subsection{Hysteresis in the Jammed Medium During Swimming}\label{sec:jam}

\begin{figure}
    \centering
    \includegraphics[width=\linewidth]{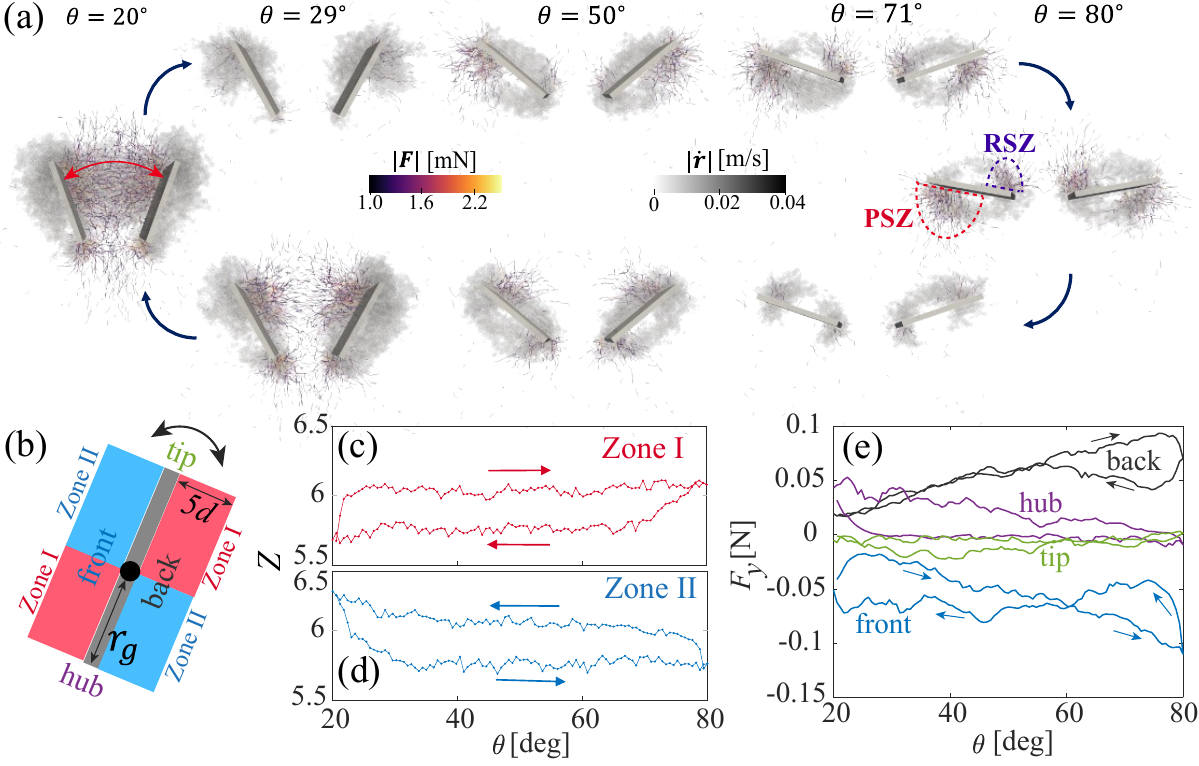}
    \vspace{-4mm}
    \caption{Force transmission and structural analyses for the base simulation case. (a) Particle velocity and inter-particle contact forces visualized during one swimming cycle. (b) Sketch for spatial decomposition of the swimmer's surroundings, \R{defining zones I and II}. (c) and (d) Average coordination number $Z$ over $\theta$ for zone I and II, respectively. (e) $y$ component of the total force exerted on different lateral surfaces of the wings.}
    \vspace{-4mm}
    \label{fig:bigfig}
\end{figure}

In Fig.~\ref{fig:bigfig}a, we visualize the inter-particle contact forces $|\boldsymbol{F}|$ and particle velocities $|\dot{\boldsymbol{r}}|$ in the base case (table \ref{tab:1}) at five different $\theta$ during both the opening and closing strokes. 
For clarity, only the top $0.1\%$ of the contact forces are shown, and only particles with velocities $|\dot{\boldsymbol{r}}|> 0.33 \omega L$ are shown. 
During each stroke, stagnant zones with percolated strong forces gradually form near the tips of the rotating wings on the leading edges that press into the medium, as in $\theta=50^\circ$ for both strokes. These contact forces form chains that are rooted from the wings and propagate into the medium, which is a signature of \R{jammed}
granular media~\citep{bi2011jamming,behringer2018physics}. 
As each stroke proceeds, a second set of force chains that resist the swimmer's translation appear near the wing hubs as seen in $\theta=71^\circ$ for opening and $\theta=29^\circ$ for closing.
We refer to these zones as resistive stagnant zones (RSZ) and the other type as propulsive stagnant zones (PSZ), which are separated by the gyration point of a wing that has zero instantaneous velocity in the direction normal to the larger lateral surface of the wings.
The distance from a gyration point to a hub is $r_g= |\dot{y}(\theta)|\sin(\theta)/\omega$, where $\dot{y}$ denotes the swimmer's translational velocity. 
\R{As particles within the stagnant zones translate almost rigidly with the swimmer wings, these zones effectively become an extension of the swimming body, and they evolve throughout each stroke. Consequently, hysteresis in the morphology and the strong force density of these stagnant zones between the opening and the closing stroke can break the translational symmetry.}

\R{To determine the jamming state of the particles near the wings,} we decompose the immediate surroundings of the wings into zones I and II, separated by $r_g(\theta)$, as depicted in Fig.~\ref{fig:bigfig}b. Both zones span the full height of the wings and extend $5d$ from the wings. 
In each zone, we average the coordination number $Z$, i.e., the number of contacts per particle, and plot
the average as a function of $\theta$ in a swimming cycle in Fig.~\ref{fig:bigfig}c and d for zones I and II, respectively.
\R{For frictional spheres, the isostatic condition of the coordination number is $(n_{\rm{dim}}+1)\le Z_{\rm{iso}}\le 2n_{\rm{dim}}$, where $n_{\rm{dim}}=3$ is the dimension~\citep{behringer2018physics},
with the specific value of $Z_{\rm{iso}}$ depending on the friction coefficient and the preparation protocol~\citep{silbert2010jamming}.}
\R{Throughout the entire cycle, the mean coordination number remains well
above the minimal threshold of $n_{\rm{dim}}+1=4$. As particles are constantly under the confining pressure from the weight of the particles above them, the particles in zones I and II are always jammed and require finite forces to be rearranged.} 

In zone I, the coordination number $Z$ increases during opening until $\theta\approx29^\circ$, corresponding to the development of force chains in the PSZ in Fig.~\ref{fig:bigfig}a. 
For $\theta>29^\circ$, $Z$ plateaus at $Z\approx6$ as the PSZ becomes fully developed and the force chain intensity remains unchanged in $50^\circ \leq \theta \leq 80^\circ$. 
After the wing rotation direction changes, $Z$ decreases towards a plateau around $Z=5.7$, corresponding to the \R{relaxation in the stagnant zones via decompression and particle rearrangement.}
Similar hysteresis in $Z$ has been reported for jammed granular systems under cyclic shear~\citep{zhang2010statistical, bi2011jamming,zhao2019shear,zhao2022ultrastable}. 

In zone II, while a similar hysteresis in $Z$ exists, a secondary increase up to $Z\approx6.5$ occurs near the end of the closing stroke, which is due to the merging of two stagnant zones as in Fig.~\ref{fig:bigfig}a from $\theta=29^\circ$ to $\theta=20^\circ$. 
\R{With $Z>2n_{\rm{dim}}$, particles in this region are likely isotropically jammed, whereas particles in zone I are more likely to experience shear jamming, evident by lower $Z$ and anisotropic force chains.}
During this period, force chains rooted from one wing land on the other, as sketched in $\theta=20^\circ$, making them effectively ``internal'' forces of the swimmer with little contribution to the propulsion. 

We then separately examine the $y$ component of the forces, $F_y$, exerted on the four lateral wing surfaces as defined in Fig.~\ref{fig:bigfig}b, which are plotted in Fig.~\ref{fig:bigfig}e as a function of $\theta$. 
While a monotonic increase can be observed in $F_y$ on the back side during opening, the front side $F_y$ plateaus near the end of closing. At this stage, a large $F_y$ arises for the hub as its surrounding region serves as the RSZ. The increasing amount of jammed particles makes it difficult for the swimmer to move in the $-y$ direction, while triggering upward particle flows as in Fig.~\ref{fig:loc}c and Appendix~\ref{app:3dvf}.

\R{We next vary the DEM configurations to systematically influence the stagnant zone formation and discuss its implication in locomotion, which is realized by simulating different swimmer geometries and a frictionless medium with $\mu=0$. }

\begin{figure}[t]
    \centering
    \includegraphics[width=\linewidth]{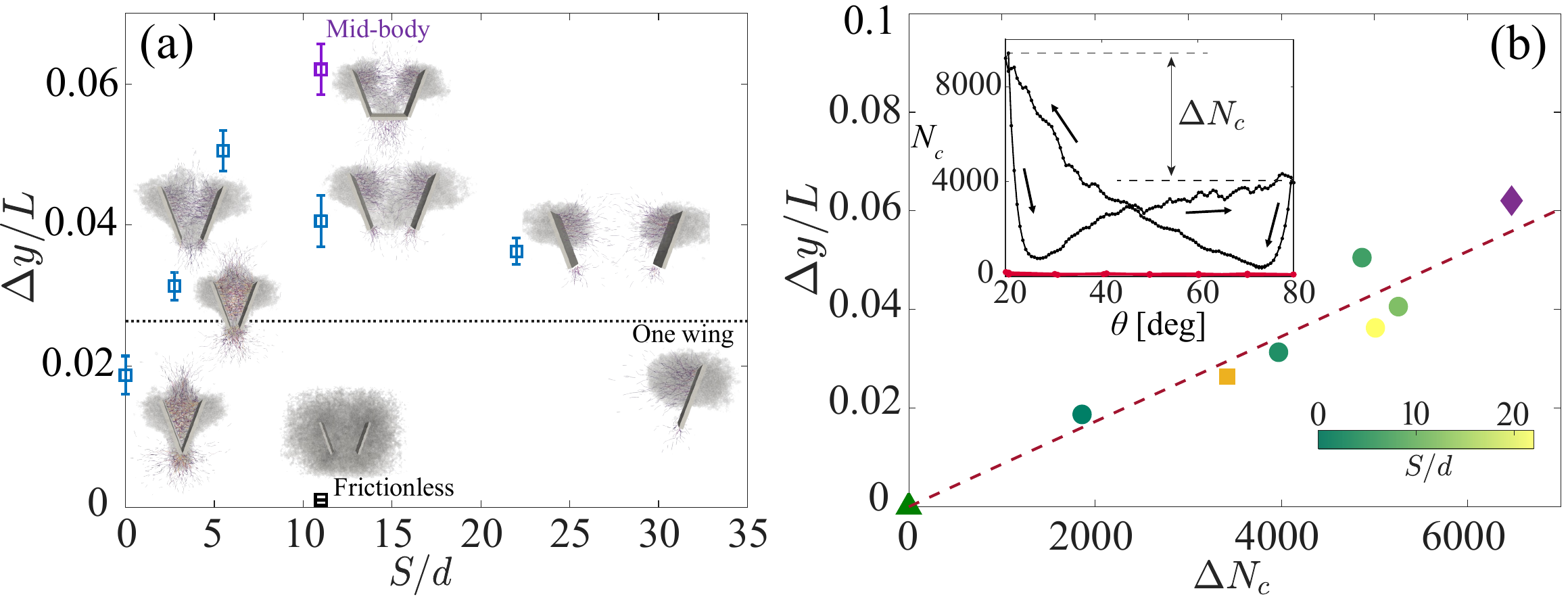}
    \vspace{-5 mm}
    \caption{The influence of jamming on net locomotion. (a) Cycle-averaged swimmer net displacement $\Delta y$ for cases with different gap widths $S$, with error bars indicating cycle-cycle fluctuations, along with the case with inert mid-body (purple) and the case with $\mu=0$ (black). The black dashed line marks the value of $\Delta y$ for the one-wing case. (b) Cycle-averaged $\Delta y$ vs. the contact number difference $\Delta N_c$. Inset shows the $N_c$ in the base case \R{(black curve with $F_c=1$ mN) and the frictionless case (red curve with $F_c=0.8$ mN)} in a swimming cycle.
    The purple diamond represents the case with inert mid-body, the orange square represents the one-wing case, and the green triangle represents the case with $\mu=0$. The red dashed line is a linear fit, $\Delta y/L = \alpha \  \Delta N_c$, where $\alpha=8.63\times10^{-6}$.}
    \label{fig:Gap}
    \vspace{-4 mm}
\end{figure}

For the geometry, we varied the wing gap width $S$ that can influence the intensity of jamming between the wings. To approximate $S=\infty$, a case with only one wing was simulated.
We also altered the swimmer geometry by inserting an inert mid-body to connect the wings to approximate a more realistic-looking swimmer.
In Fig.~\ref{fig:Gap}a, we show how the swimmer's normalized net displacement per cycle, ${\Delta y/L}$, depends on the varied factors. 
As the gap width increases, ${\Delta y/L}$ rises until $S/d \approx 6$, then decreases for ${S/d > 6}$ towards the value for $S/d = \infty$ (the black dashed line). This trend resembles the cooperative effects reported for two closely-spaced intruders in granular media, where both the drag force \citep{agarwal2021efficacy} and total work \citep{pravin2021effect} peak at a separation of approximately $4d$.
Similar to the result with inert rods in Fig.~\ref{fig:loc}a, by filling this gap with an inert mid-body, a significant increase in $\Delta y/L$ occurs, accompanied by a stagnant zone (RSZ) with more intense force chains behind the mid-body.

In the simulation with $\mu=0$, locomotion vanishes \R{(Fig.~\ref{fig:loc}a)}, coinciding with the disappearance of the stagnant zones as seen in the snapshot inside Fig.~\ref{fig:Gap}a.
\R{The initial packing fraction is $\phi_0=0.64$ and coordination number is $Z=6.8$, which is similar to the simulated value by~\cite{silbert2002statistics}, indicating that the frictionless particles are still jammed under gravitational forces.}
\R{Different from the frictional cases shown in Fig.~\ref{fig:bigfig}a and Fig.~\ref{fig:Gap}a, a larger number of particles are mobilized uniformly around the swimmer, avoiding the hysteretic buildup of strong forces. In other words, the lack of interparticle friction facilitates local rearrangements that prevent the accumulation of stagnant zones near the wings.}


\R{To quantify the ``intensity'' of stagnant zones induced by the swimmer, we define a strong contact number, $N_c$, representing the number of particle contacts where the force magnitude exceeds a threshold $F_c$. 
This threshold is set to the top 0.1\% of the contact force distribution across the entire system, which roughly includes all the visualized strong forces in the stagnant zone in Fig.~\ref{fig:bigfig}a. }
\R{For the base case with the frictional medium, we extract a threshold of $F_c \approx 1.0$\,mN at the time instant $\theta=20^\circ$, and we apply this value to all other frictional cases. For the frictionless case, we obtained $F_c \approx 0.8$\,mN.} 
\RR{The threshold value coincides with the largest contact forces at the bottom of the container, due to gravity, as estimated by $\rho \phi_0 g H (\pi d^2/4)\approx 1$\,mN, with $H$ being the bed height.
In this sense, forces exceeding this threshold are considered ``strong'' relative to the full granular medium. 
However, $N_c$ should represent the intensity and localization of the high-force contact network associated with stagnant-zone formation.
Thus, this definition requires us to sample near the swimmer, so that forces at the bottom of the container are not included, as they do not contribute to the locomotion of the swimmer. Therefore, we only consider contacts that are within a region, centered around the swimmer, with a size of $55d \times 40d \times 25d$.
}
\RR{The robustness of this definition is examined in Appendix~\ref{app:robustness}, where we compare force distributions, normalized high-force tails, and various versions of the threshold.}

An example of $N_c$ vs $\theta$ in a swimming cycle is shown in the inset of Fig.~\ref{fig:Gap}b, showing a broken symmetry in the force transmission between the opening and closing strokes, agreeing with the structural signal based on $Z$ in Fig.~\ref{fig:bigfig}c and d. \R{For the frictionless case, however, a very low number of strong contacts exists near the swimmer, as most of the top 0.1\% forces reside at the bottom, again showing a lack of stagnant zones.}

As argued earlier, the excess number of strong contacts towards the end of the closing stroke can limit the swimmer's translation. Therefore, we quantify the degree of \RR{stagnant-zone-induced} symmetry breaking by calculating the difference of $N_c$ between the fully closed and fully open states, yielding 
\begin{equation}
\Delta N_c = N_c(\theta_{\mathrm{c}})-N_c(\theta_{\mathrm{o}}). 
    \label{eq:nc}
\end{equation}

\noindent Figure~\ref{fig:Gap}b shows the net displacement ${\Delta y/L}$ with respect to $\Delta N_c$, revealing a linear relation between these two physical quantities across all cases with varied swimmer geometry and friction. This quantitatively demonstrates how stagnant-zone-induced effects in jammed media break the symmetry between the two reciprocal strokes of the scallop swimmer and enable net locomotion in the $+y$ direction. 


\RR{To rationalize the relation between $\Delta N_c$ and ${\Delta y/L}$, we adopt an argument based on ``kinematic compatibility''~\citep{xiao2024locomotion}. 
For each stroke, the swimmer's wings sweep across a volume that is $L^3(\theta_{\mathrm{o}}-\theta_{\mathrm{c}})$. If the particles rearrange freely, this volume can be compensated by the swimmer translating a distance $y_s$ accordingly, sweeping a volume that is approximately $2L^2\sin(\theta_{\mathrm{m}})y_s$, with $\theta_{\mathrm{m}}=(\theta_{\mathrm{o}}+\theta_{\mathrm{c}})/2$. Equating the two volumes, we obtain $y_s\approx0.7L$, which is comparable to the observed $y_s$ in Fig.~\ref{fig:loc}a, especially for the frictionless case that has the largest $y_s$ per stroke.
When particles can easily rearrange, \textit{e.g.}, with $\theta=50^\circ$ in Fig.~\ref{fig:flow}, the vortex-like flow field develops, which corresponds to the flow of particles from Zone I towards Zone II (Fig.~\ref{fig:bigfig}b) during opening and vice versa.}

\RR{For swimming in a jammed, solid-like medium with frictional particles, a subset of particles, characterized by $N_c$, develop strong resistance and do not rearrange to accommodate the swimmer's translational motion.
This effect is most pronounced at small $\theta$ during the closing stroke when $N_c$ peaks. For example, at $\theta=29^\circ$, the vortex-like particle transport breaks down on the \emph{inner side} of the wings. Particles in Zone II (purple star in Fig.~\ref{fig:flow}) are highly crowded, as indicated by the converging in-plane velocity and the concentration of strong forces (Fig.~\ref{fig:bigfig}a), and they cannot effectively flow towards Zone I (white star). Without these particles replenishing Zone I, the hubs moving away from Zone I (in the $-y$ direction) would tend to create a ``vacuum'' in the zone. This is kinematically incompatible in a dense granular medium, and therefore the swimmer translation $y_s$ is hampered. Instead of pumping particles in the $y$ direction to propel itself, the swimmer lifts particles vertically in the $z$ direction, as shown in Appendix~\ref{app:3dvf}.}

\RR{A simplified way to account for the effect of $N_c$ is to consider the corresponding characteristic volume, $2N_cd^3/\bar{Z}$, as a ``negative'' volume to be subtracted from the wing rotation-swept volume that is available for the swimmer translation. Here, $\bar{Z}$ is a nominal coordination number, introduced to convert the contact count to the particle count. This viewpoint is equivalent to the viewpoint that the swimmer is effectively enlarged by $2N_cd^3/\bar{Z}$ and needs to move together with these stagnant particles.
The characteristic volume balance for each stroke is then $L^3-2N_cd^3/\bar{Z}=2L^2y_s$, noting that the $\theta$-related factors are dropped and the $N_c$ used here is measured at the end of each stroke, consistent with the calculation of $\Delta N_c$. Taking the difference between the two strokes and estimating $\bar{Z}=6$, we recover the linear relation, $\Delta y/L=[d^3/(6L^3)]\Delta N_c$, noting that the rotation-swept volume cancels out due to reciprocity. Here, the proportionality factor, $d^3/(6L^3)=5\times10^{-5}$, is within an order of magnitude of the fitted constant, $\alpha=8.63\times10^{-6}$, from Fig.~\ref{fig:Gap}b, which is reasonable given the simplicity of the argument. For instance, not all $N_c$ may contribute to such ``negative'' volume, and accounting for that would lower the proportionality factor. That said, this linear relation may break down at shallower burial depths, where particles are less confined and the free surface is free to deform plastically, altering the history-dependent response of the medium.
}

\subsection{Role of Swimmer Inertia in Reciprocal Swimming}\label{sec:iner}

A second contribution to the locomotion, associated with the swimmer's inertia, emerges with increasing swimmer mass $m_s$ and decreasing swimming cycle period $T$.
To quantify the inertial effect, we consider a coasting time, 
\begin{equation}
    T_{c} = \frac{v}{F_D/m_s} = \frac{L/T} {P_sL^2/m_s}=\frac{m_s} {P_sLT},
    \label{eq:15}
\end{equation}

\noindent where $L/T$ sets a characteristic swimmer translation speed that is comparable with the swimmer's speed near the end of a stroke, i.e., the slope of the curves in Fig.~\ref{fig:loc}a. 
We consider a characteristic acceleration, $P_sL^2/m_s$, where $P_s=\rho\phi_0 gH_0$ is a hydrostatic-like pressure, and $H_0$ is the swimmer depth with respect to the free surface. In this case, \RR{$P_sL^2$ serves as a normalization factor, which should be on the same order of magnitude as the drag force.
This term comes from the rate-independent term in a previously established drag law in granular intrusion, $F_D = a_1 P_s A + a_2\rho \phi_0 A v^2$, where $A\sim L^2$ is the projected frontal area, $v\sim L/T$ is the intruder speed, and 
$a_1, a_2$ are material and geometric prefactors~\citep{albert1999slow,katsuragi2007unified,clark2012particle,brzinski2013depth,hilton2013drag,takehara2014friction}. 
The Froude number $\mathrm{Fr}= v/\sqrt{gH_0} = {v}/{\sqrt{P_s/(\rho \phi_0)}} \lesssim 0.54$ across the entire parameter sweep of table~\ref{tab:1}, so the rate-independent term dominates and $F_D\approx a_1 P_s L^2$. 
Thus, the use of $P_sL^2$ as a normalization factor is reasonable and convenient, as it remains a constant across all our simulation configurations.}
\R{Under this definition, $T_c$ reflects a characteristic time for a moving swimmer to ``coast down'' until its translational motion ceases due to the medium's drag \citep{hubert2021scallop}, were it to stop actuating.}

\begin{figure}[t]
    \centering
    \includegraphics[width=\linewidth]{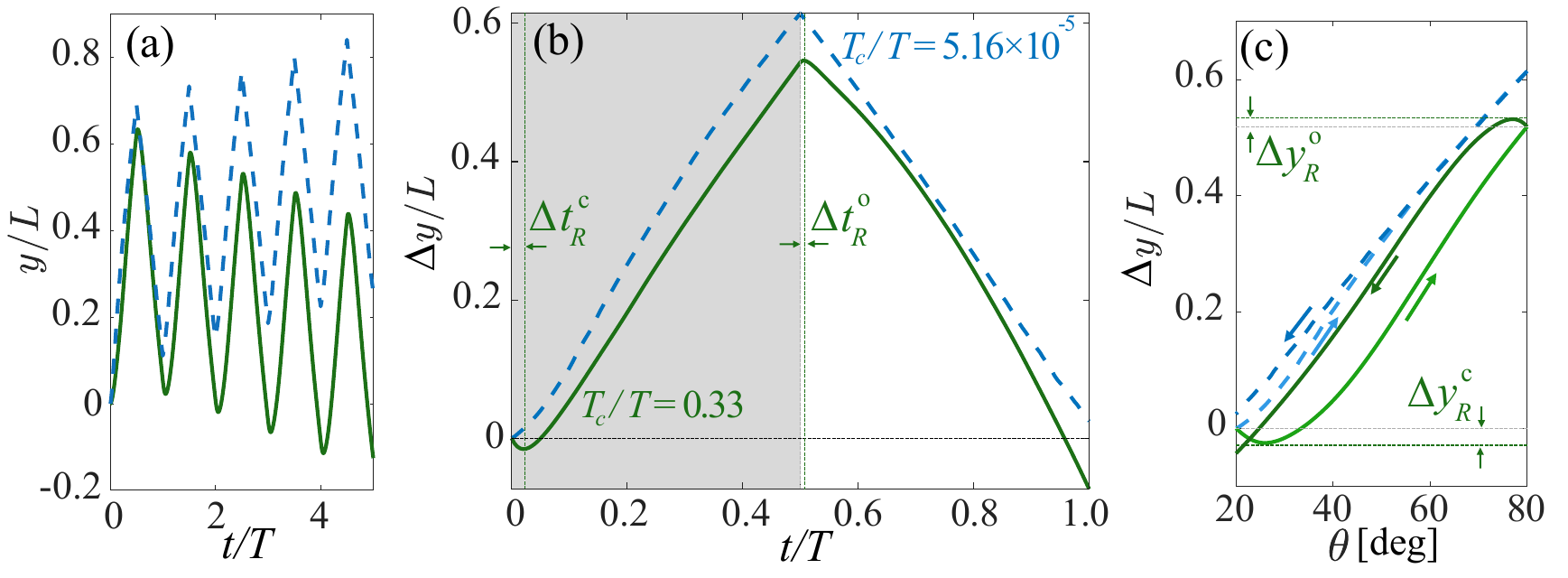}
    \caption{Locomotion in the dynamic regime. (a) The normalized swimmer displacement $y$ vs time in quasi-static (dashed blue curve) and dynamic (solid green curve) regimes. The quasi-static case is with $m_s=0.60 \text{ g}$, $T=2.0 \text{ s}$, and $T_c/T=5.16\times10^{-5}$, and the dynamic case has $m_s=2.38 \text{ g}$, $T=0.05 \text{ s}$, and $T_c/T=0.33$.
    \R{(b) Normalized $\Delta y$ vs $t/T$ for one swimming cycle for the quasi-static (dashed curve) and dynamic (solid curve) cases. The gray region marks the opening stroke, and the vertical dashed lines mark the time instants at which the direction of the swimmer's translational motion switches.}
    (c) Normalized $\Delta y$ vs $\theta$ for the two cases, with lighter colors representing opening and darker colors representing closing. \R{The horizontal dashed lines mark the $\Delta y$ at which the direction of the swimmer's translational motion switches.}
    }
    \label{fig:InertiaSchem}
\end{figure}

Figure~\ref{fig:InertiaSchem}a shows the swimmer displacement over time for a simulation in the dynamic regime
with $T_c/T=0.33$, along with the base case with $T_c/T=5.16\times10^{-5}$.
Contrary to the forward ($+y$) quasi-static locomotion~\citep{xiao2024locomotion}, a persistent locomotion occurs in the $-y$ direction, indicating a distinct mechanism in this regime. This direction agrees with the theoretical prediction of a scallop-like swimmer with non-negligible inertia in Newtonian fluids at low Reynolds numbers~\citep{gonzalez2009reciprocal}, despite the differences in the nature of forcing and problem setup.

We scrutinize $\Delta y/L$ within a single swimming cycle, as shown in \R{Fig.~\ref{fig:InertiaSchem}b}.
\R{As the swimmer never stops actuating, there is no strict coasting. Instead, we see}
a retardation process characterized by a time interval, $\Delta t^i_R$, measuring the delay between the instant when the wing rotation reverses and the instant when the swimmer's translation direction reverses \R{(marked by the vertical dashed lines in Fig.~\ref{fig:InertiaSchem}b)}, which can be seen after both the opening ($i=\mathrm{o}$) and closing ($i=\mathrm{c}$) strokes. 
Likewise, we measured the corresponding retardation distance, $\Delta y^i_R$, which is made clear in Fig.~\ref{fig:InertiaSchem}c that shows $\Delta y$ against the wing angle $\theta$ in a cycle.
\R{Both results show that for $T_c/T=0.33$, significant retardation occurs, confirming that the case is in the dynamic regime. For $T_c/T=5.16\times10^{-5}$, we see that both $\Delta t^i_R$ and $\Delta y^i_R$ are negligible, indicating a quasi-static regime where the swimmer inertia is insignificant.}



\begin{figure}[t]
    \centering
    \includegraphics[width=\linewidth]{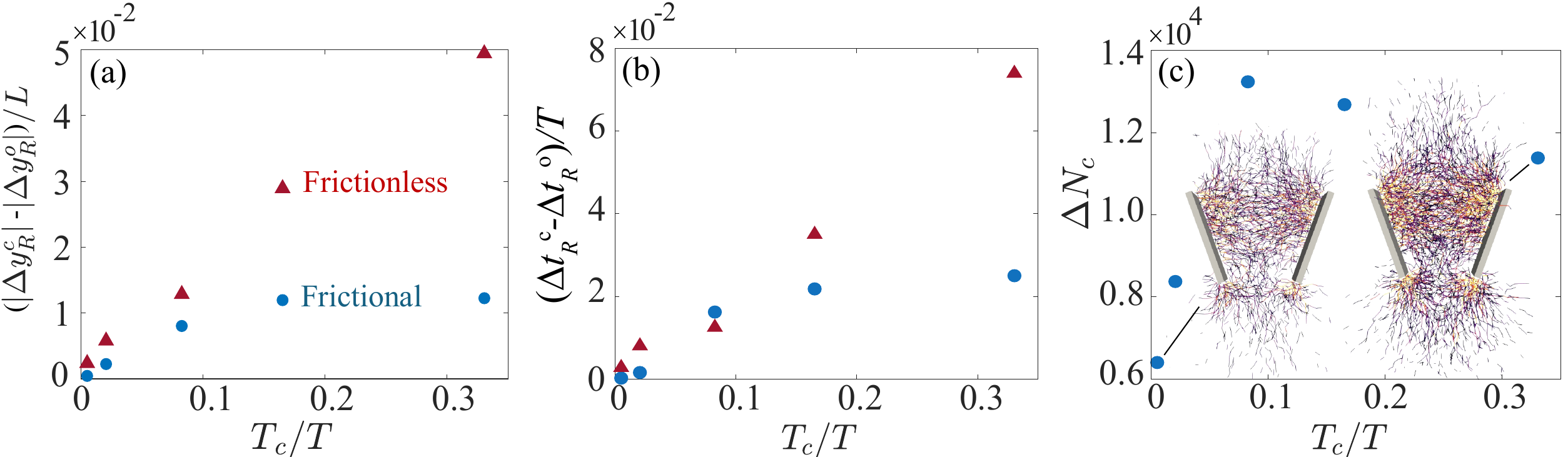}
    \caption{The dependence of $\Delta y^i_R$ and $\Delta t^i_R$ on different $T_c/T$ in both frictional and frictionless media by varying $T$ and $m_s$ according to table~\ref{tab:1}. (a) The difference in retardation distance between the two strokes vs the coasting time $T_c/T$ in various cases with frictional (blue circle) and frictionless (red triangle) media. 
    \R{(b) is the same as (a) but for the difference in the retardation time. (c) The contact number difference $\Delta N_c$ vs the coasting time $T_c/T$ for the cases in the frictional medium. 
    The two snapshots of the swimmer with force chains visualized correspond to the two cases with $T_c/T=0.005$ and $T_c/T=0.33$, which show the intensity of the contact force network in a similar color scheme as Fig.~\ref{fig:bigfig}a.}
    }
    \label{fig:Ret2}
\end{figure}

\R{For $T_c/T=0.33$, the retardation after the closing stroke is more prolonged in comparison to the retardation after the opening stroke, yielding $\Delta t^c_R>\Delta t^o_R$ and $|\Delta y^c_R|>|\Delta y^o_R|$.}
To further understand this difference, a parametric study was performed, varying $T$ and $m_s$ to determine the dependence of $\Delta y^i_R$ and $\Delta t^i_R$ on different $T_c/T$ in both frictional and frictionless media. 
\R{The absence of friction eliminates the formation of stagnant zones as discussed in Sec.~\ref{sec:jam}, which helps highlight the role of inertial coasting.}
We show $|\Delta y^c_R|-|\Delta y^o_R|$, in Fig.~\ref{fig:Ret2}a, and $\Delta t^c_R-\Delta t^o_R$ in Fig.~\ref{fig:Ret2}b, both confirming that the retardation is more prolonged after the end of the closing stroke, and the differences increase with $T_{c}/T$. 

The difference in the retardation behavior is reminiscent of the result in \cite{hubert2021scallop} that a floating dumbbell swimmer can propel itself provided that the respective coasting times for its reciprocal strokes are different.
\R{Likewise, the symmetry-breaking mechanism in the inertial regime of the scallop swimmer is related to this difference in the retardation behavior.
In the reciprocal strokes of opening and closing, the swimmer gains similar peak velocity, as reflected by the slope of $\Delta y$ in Fig.~\ref{fig:InertiaSchem}c near $\theta=50^\circ$, noting that $\theta$ is linear with $t$.}
However, near the end of each stroke, including the beginning period of the subsequent stroke, the resistive drag force that opposes the swimmer's translation is different between opening and closing.   
This force should be proportional to the swimmer's exposed area in the $x$-$z$ plane, $2L^2\sin(\theta)$, which is larger after opening than closing. This results in the more prolonged retardation after the closing stroke, hence the preferred backward ($-y$) locomotion direction.

\begin{figure}[t]
    \centering
    \includegraphics[width=\linewidth]{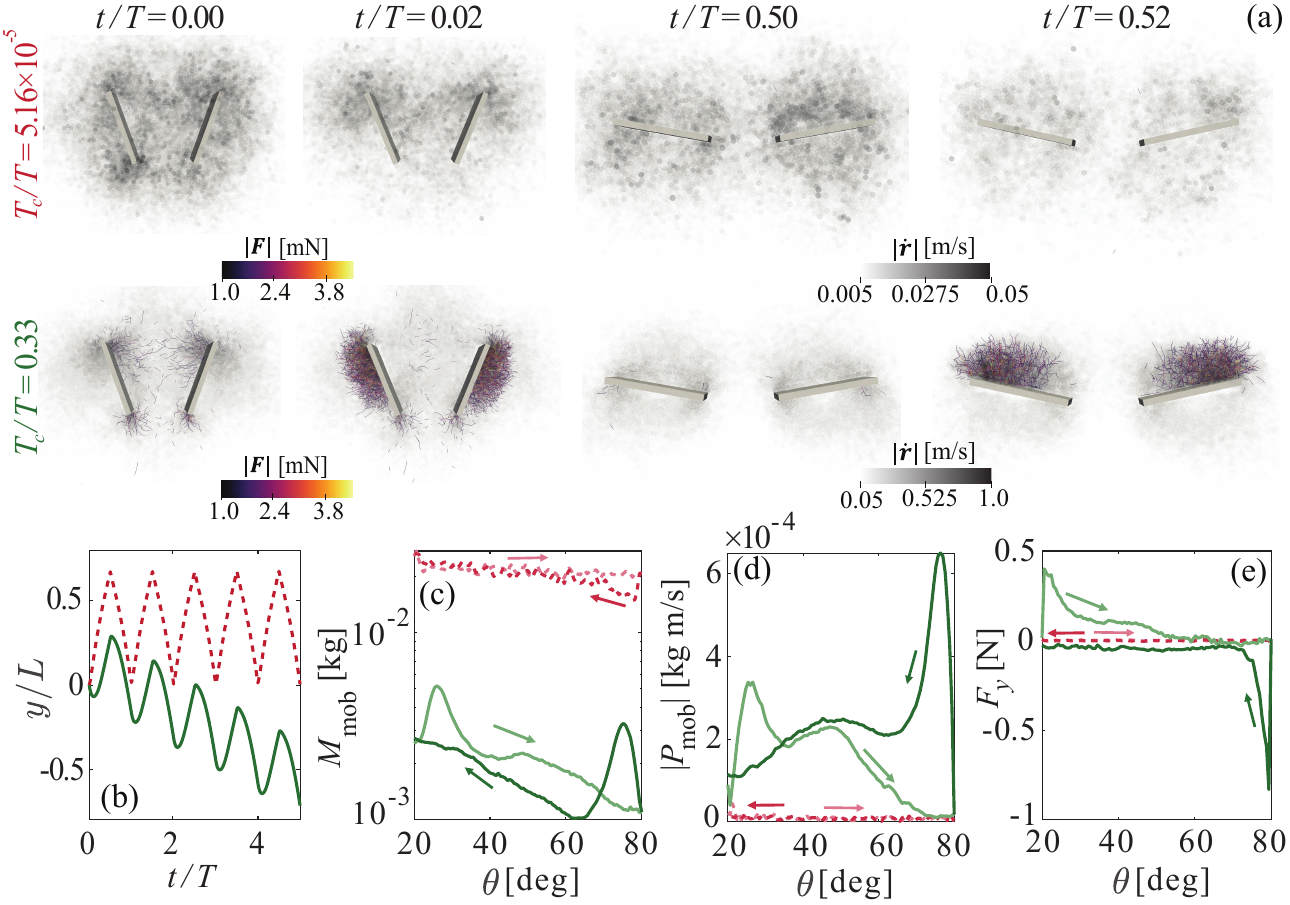}
    \caption{\RR{Particle-scale analyses of the frictionless granular medium at two different $T_c$. (a) Visualized particle velocity and inter-particle contact force at the end of each stroke and at $0.02 \,T$ into the subsequent stroke. The first row is for a swimmer with $m_s=0.60 \text{ g}$, $T=2.0 \text{ s}$, and $T_c/T=5.16\times10^{-5}$, and the second row is for $m_s=2.38 \text{ g}$, $T=0.05 \text{ s}$, and $T_c/T=0.33$.
    Only the top 0.1\% of the contact forces and the particles with $|\boldsymbol{\dot r}|>0.33\omega L$ are shown.
    (b) Swimmer displacement over time.
    (c) Mass of mobilized particles, whose velocity is larger than $0.33 \omega L$ during a swimming cycle. (d) Momentum magnitude of the mobilized particles. (e) $y$ component of the medium's resistive force applied to the swimmer. In (b--e), the red dashed curves correspond to $T_c/T=5.16\times10^{-5}$ and the green solid ones to $T_c/T=0.33$. The lighter and darker curves correspond to the opening and closing strokes, respectively.}
    }
    \label{fig:InerMic}
\end{figure}

\RR{To isolate the contribution of inertia to the backward locomotion, we further examine particle-scale features in the frictionless medium, which suppresses the friction-induced hysteresis as in Sec.~\ref{sec:jam}. We compare a quasi-static case with $T_c/T = 5.16\times10^{-5}$ and a dynamic one with $T_c/T = 0.33$, visualizing the inter-particle contact forces and particle velocities at the end of each stroke and at $0.02 \,T$ into the subsequent stroke, which is within the retardation process for the dynamic case. 
Without particle friction, the backward locomotion displacement over time is larger for $T_c/T = 0.33$ than that with frictional particles, which can be seen by comparing Fig.~\ref{fig:InerMic}b to Fig.~\ref{fig:InertiaSchem}a.
We also track the total mass, $M_\mathrm{mob}$, of mobilized particles (with $|\boldsymbol{\dot r}|>0.33\omega L$), the corresponding total particle momentum magnitude $P_{mob}(t) = \sum_j m_j|\dot{\boldsymbol{r}}_j|$, and the $y$ component of total resistive force on the swimmer, which are shown in Fig.~\ref{fig:InerMic}c--e, respectively.}

\RR{Between the two cases, major differences are found in the particle-scale response to the wing-rotation reversal, as shown in Fig.~\ref{fig:InerMic}a. For the quasi-static case, almost no concentrated, strong contact network forms at $t/T=\{0.0, 0.5\}$ and at $t/T=\{0.02,0.52\}$. 
A wide range of particles are in motion and the velocity of the mobilized particles is smaller at $t/T=\{0.02,0.52\}$ compared to those at $\{0.0, 0.5\}$. 
As $T_c/T$ is small, particles have enough time to reorganize and dissipate the transferred momentum from the swimmer, as confirmed in Fig.~\ref{fig:InerMic}c, showing a persistently large mobilized mass over a cycle. 
Also, the transferred momentum and resistive force in this regime are negligible, as shown in Fig.~\ref{fig:InerMic}d,e, confirming that no granular inertia is carried across strokes.}

\RR{For the dynamic case, however, a strong, concentrated contact network emerges after $t/T=\{0.0, 0.5\}$ and further intensifies at $t/T=\{0.02,0.52\}$.
This is a consequence of the swimmer coasting, i.e., the wings both rotating into and translating into the corresponding particles at high speeds. As a result, particles do not have enough time to rearrange to dissipate the transferred momentum throughout the medium and instead develop large contact forces.
In fact, far fewer particles exceed the $0.33\omega L$ speed threshold as shown qualitatively in Fig.~\ref{fig:InerMic}a and quantitatively in Fig.~\ref{fig:InerMic}c.
Similarly, we observe an increase in momentum and resistive force at $\theta$ that corresponds to $t/T=\{0.02,0.52\}$ in Fig.~\ref{fig:InerMic}d,e. 
As a result, all three quantities exhibit peaks during swimmer coasting. 
More importantly, these peaks are asymmetric between the two strokes, with peaks following the opening stroke being larger in magnitude compared to those following the closing stroke, which, again, is due to the geometric asymmetry of the wings. At $\theta\approx20^\circ$, the concentrated strong forces result in mostly lateral forces (in the $x$ direction) on the wings, whereas at $\theta\approx80^\circ$, $|F_y|$ is larger, which impairs the coasting motion as discussed earlier.} 

\RR{Even without friction, particles with concentrated strong forces (Fig.~\ref{fig:InerMic}a) also constitute stagnant zones due to the lack of relaxation, i.e., their jammed packing structure remains ``stagnant.'' However, key differences exist between these zones that are rate-dependent and those friction-dependent zones as discussed in Sec.~\ref{sec:jam}. The rate-dependent zones appear at the beginning of each stroke and resist coasting motion in the direction set by the previous stroke, whereas the friction-dependent zones become most prominent at the end of each stroke and resist motion in the direction set by the current stroke. Hence, the respective propulsion mechanisms, related to the symmetry-breaking of these zones, result in opposite propulsion directions. Of course, for dynamic swimming in a frictional medium (Fig.~\ref{fig:InertiaSchem}), both mechanisms coexist, and we next seek a more unified description.}



\subsection{Coexistence and Unified Scaling of the Two Mechanisms}\label{sec:unif}

\RR{Having characterized the jamming mechanism in Sec.~\ref{sec:jam} and the inertial mechanism in Sec.~\ref{sec:iner} in isolation, we now examine how the two mechanisms coexist based on the simulations for}
Fig.~\ref{fig:Ret2}. The increase of $|\Delta y_R^c|-|\Delta y_R^o|$ and $\Delta t_R^c-\Delta t_R^o$ with $T_c$ is linear in the frictionless cases, while it saturates in the frictional cases at large $T_c$. 
\R{The saturation is likely due to the competition from the stagnant zone formation due to jamming, as discussed in Sec.~\ref{sec:jam}. For example, during retardation after the closing stroke, the rotating wing is pressing towards the $-y$ direction while the swimmer is also moving towards the $-y$ direction under its inertia. This may significantly compress the particles in the backside of the swimmer, inducing a large resistive stagnant zone (RSZ) that hinders the retardation distance and results in the saturation in Fig.~\ref{fig:Ret2}a and b. From example snapshots inside Fig.~\ref{fig:Ret2}c, we see a larger RSZ for the case with higher $T_c/T$.} 

\RR{Therefore, the jamming-based and inertia-based mechanisms are not mutually exclusive regimes; they coexist and contribute to locomotion with different parameter dependencies. The jamming-based contribution is governed by the friction coefficient $\mu$ and the hysteretic nature of the medium, which manifests as asymmetric stagnant zones towards the end of the two strokes as captured by $\Delta N_c$. The inertial contribution is governed by the ratio of the coasting time to the swimming period, $T_c/T$, which is more relevant to the swimmer dynamics at the beginning of the two strokes. 
It is also worth noting that the two mechanisms mentioned in this study appear as competing contributions in the net locomotion at high $T_c/T$. While the jamming-based mechanism favors the forward motion, the inertia-based mechanism favors the backward motion.}

\begin{figure}[t]
    \centering
    \includegraphics[width=\linewidth]{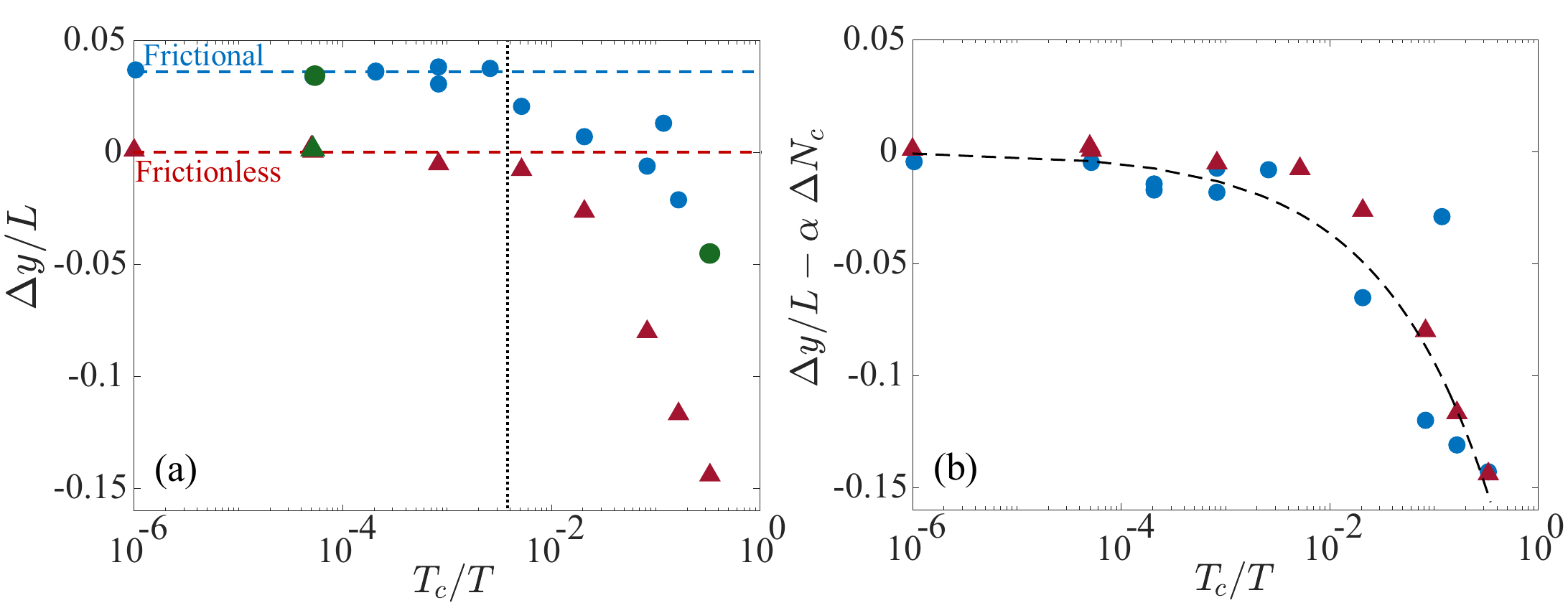}
    \caption{Quantified relations between swimmer locomotion and inertial coasting. (a) Normalized swimmer displacement vs the coasting time in the frictional medium (blue circles) and the frictionless medium (red triangles). Green circle markers show the simulation cases demonstrated in Fig.~\ref{fig:InertiaSchem} and the green triangle marker shows the case in the frictionless medium demonstrated in Fig.~\ref{fig:loc}a. (b) The same plot but with $\Delta y/L$ offset by the jamming contribution $\alpha\Delta N_c$. \RR{The dashed curve is the fitted Eq.~\eqref{eq:16}.}
    }
    \vspace{-1 mm}
    \label{fig:Inertia}
\end{figure}

\R{We then seek to construct a unified description considering both mechanisms.} 
In Fig.~\ref{fig:Inertia}a, we relate the net swimmer locomotion to the coasting time $T_c/T$, which reflects both the significance of the swimmer's inertia and the difference between the coasting after the closing and opening strokes.
In the frictional cases, a plateau exists at low $T_c/T$, showing a quasi-static regime as discussed in Sec.~\ref{sec:jam}. At $T_c/T\approx 4 \times 10^{-3}$, a transition to a dynamic regime occurs, showing a decrease in forward locomotion or even the presence of backward locomotion. In the frictionless cases, the same plateau exists in the quasi-static regime, but without forward locomotion. A transition occurs at a similar $T_c/T$ as in the frictional cases, after which the swimmer shows a significant increase in backward locomotion. 
The difference in $\Delta y/L$ between the frictional and the frictionless cases is related to $\Delta N_c$ as shown in Fig. \ref{fig:Gap}b. 
After subtracting $\alpha\Delta N_c$ from $\Delta y/L$, the frictional and frictionless results collapse in Fig.~\ref{fig:Inertia}b, where a master power-law curve fits well to all data, as 

\vspace{-3 mm}
\begin{equation}
    \Delta y / L = \alpha \Delta N_c - \beta \ (T_c/T)^{n}
    \label{eq:16},
\end{equation}

\noindent \AN{with the fitted $\beta=0.24$ and $n=0.44$,} highlighting the respective roles of jamming and swimmer inertia in locomotion. 
\R{The collapse of data suggests that the locomotion of the scallop swimmer can be decomposed into a contribution from jammed stagnant zone formation, characterized by $N_c$, and a contribution from the inertial coasting, characterized by $T_c/T$.}
We note that Eq.~\eqref{eq:16} should not be extrapolated to $T_c/T=\infty$. As the swimmer keeps increasing its flapping frequency, the granular medium can be strongly agitated and transition to a gaseous behavior, which can possibly lead to a decrease in $|\Delta y|$. Similarly, if the swimmer's mass $m_s$ keeps increasing, the swimmer may not be able to gain enough momentum from a stroke, also resulting in decreased $|\Delta y|$. 

\vspace{-3 mm}
\section{Summary}
For the scallop-inspired granular swimmer, reciprocal swimming is possible via two distinct mechanisms. The first is related to jamming in granular matter. The timescale (or wing rotation $\theta$) over which jammed stagnant zones develop overlaps with the swimming cycle period (or $\theta_{\text{o}}-\theta_{\text{c}}$), inducing hysteresis in the medium response during swimming.
The scallop-inspired geometry results in stronger jamming effects towards the end of the closing stroke, further breaking the symmetry in both the structure and the force transmission in the medium between opening and closing. The resulting forward locomotion has a net displacement that is linear to the difference in the number of strong particle contacts between closing and opening. 
In the dynamic regime where the swimmer's inertia becomes important, the second swimming mechanism relies on the difference between the swimmer's coasting times during closing and opening.
The retardation time and distance at the end of closing are larger than those at the end of opening, resulting in an overall backward locomotion, with the net displacement related to the normalized coasting time, $T_c/T$.

\RR{From a mechanics perspective, the presented granular swimming problem combines the complexity of the elasto-plastic deformation of a jammed solid-like medium and the viscoelastic flow of a complex fluid.
The understanding of how to navigate such complex media has essential implications in granular locomotion.} The jamming-based mechanism in the quasi-static regime shows potential for designing the geometry of the swimmer to achieve optimal locomotion. Specifically, in Fig.~\ref{fig:loc}a and Fig.~\ref{fig:Gap}a, we see that the swimmer can generate larger locomotion if its body has an inert component. As for the inertia-based mechanism, although we currently disabled vertical translation of the swimmer, it is possible that a truly free swimmer with a density higher than the medium can be sustained due to the solid-like nature of granular matter or under micro-gravity.
Future research could explore the effects of varying the geometry and mass of the swimmers and with more degrees of freedom unlocked. 

\begin{appen}
\section{Three-dimensional velocity field and force-chain connectivity}\label{app:3dvf}

\begin{figure}[h!]
    \centering
    \includegraphics[width=\linewidth]{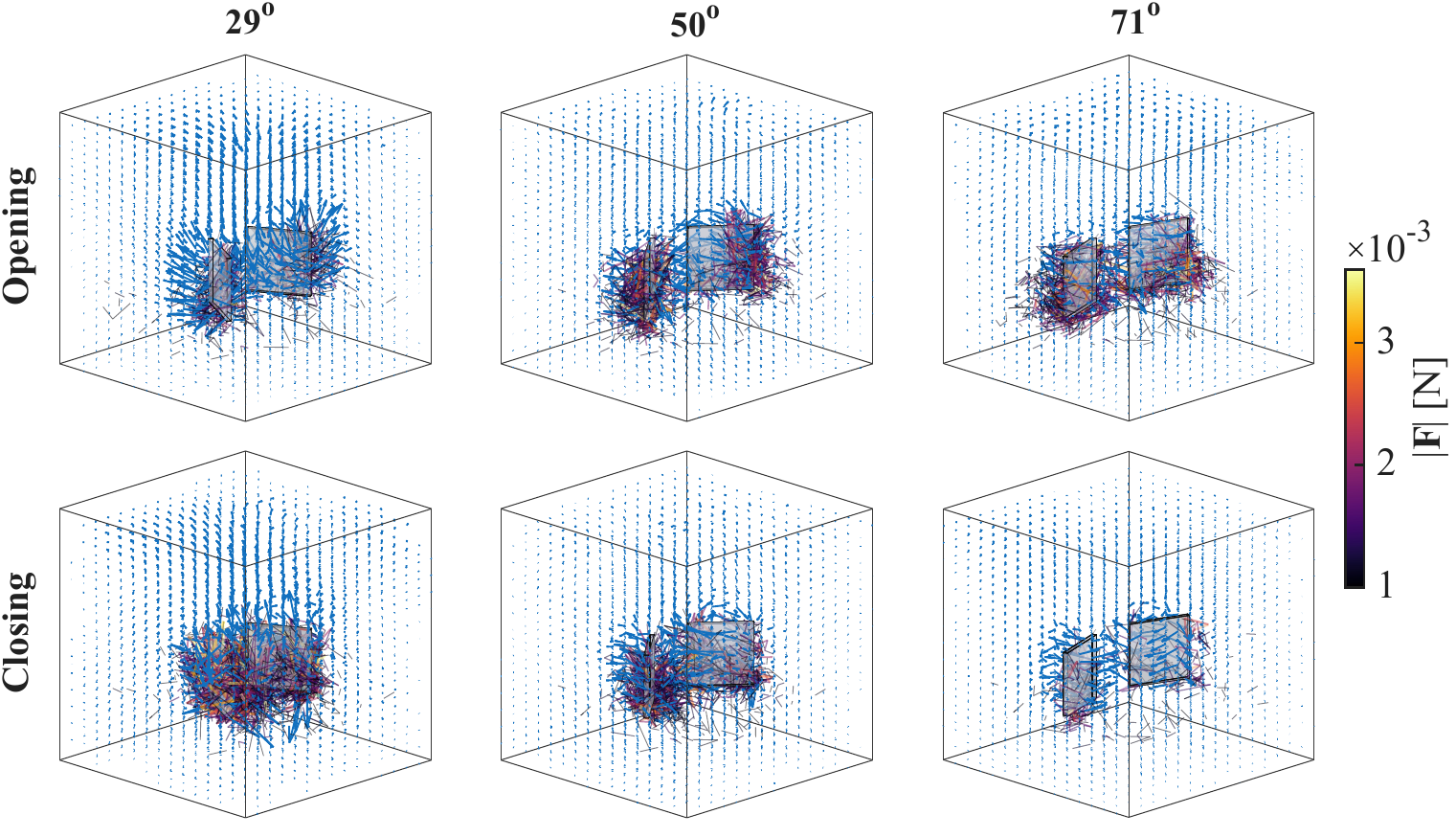}
    \caption{\RR{Three-dimensional visualization of particle velocities (vectors) and strong inter-particle contact forces (semi-transparent colored lines connecting the contacting particle centers) near the swimmer for the frictional base case. The colormap shows the magnitude of the contact force and the arrow length indicates the magnitude of the velocity. Top and bottom rows show opening and closing strokes, respectively; columns correspond to three wing angles $\theta=29^\circ$, $50^\circ$, and $71^\circ$.} 
    }
    \label{fig:3dvf}
\end{figure}

\RR{We visualize the 3D velocity field and strong contact networks in the whole domain in Fig.~\ref{fig:3dvf}. The velocity field is obtained following the same coarse-grained method as in Sec.~\ref{sec:val}, and is shown as 3D vectors, whose length indicates the magnitude of velocity. Meanwhile, contact forces are represented as semi-transparent lines connecting the contacting particles' centers, whose color indicates the magnitude of the contact force. It should be noted that we only visualize contacts that have force magnitude larger than the strong threshold $F_c$ and are within the region of interest.}

The 3D fields in Fig.~\ref{fig:3dvf} corroborate the 2D mid-plane views in Figs.~\ref{fig:loc},~\ref{fig:flow},~and~\ref{fig:bigfig}, while revealing features that the 2D slices cannot capture. At intermediate and large wing angles, $\theta=50^\circ,71^\circ$, the particle motion remains largely confined to the $x-y$ plane, whereas at $\theta = 29^\circ$ a pronounced $z-$component of velocity appears. The rotation of the wings within the narrow gap between them forces particles to move out of the $x-y$ plane. Gravity, which imposes a vertical pressure gradient in the medium, biases this out-of-plane motion, driving an upward convection of the medium during closing and a downward convection during opening stroke. This strong out-of-plane displacement at $\theta = 29^\circ$ is consistent with Fig.~\ref{fig:flow}, where $\dot{\gamma}$ is most pronounced. Figure~\ref{fig:3dvf} also shows the strong-contact network extending from the leading edges of the wings into the depth of the packing, revealing the 3D shape of the stagnant zones. Moreover, the intensity of the strong contacts differs between opening and closing, most visibly at $\theta=29^\circ$ and $\theta=71^\circ$, reflecting the history-dependence and symmetry-breaking of the frictional medium in response to the reciprocal swimming motion.

\section{Force statistics and robustness of the force threshold $F_c$}\label{app:robustness}

\begin{figure}[t]
    \centering
    \includegraphics[width=0.9\linewidth]{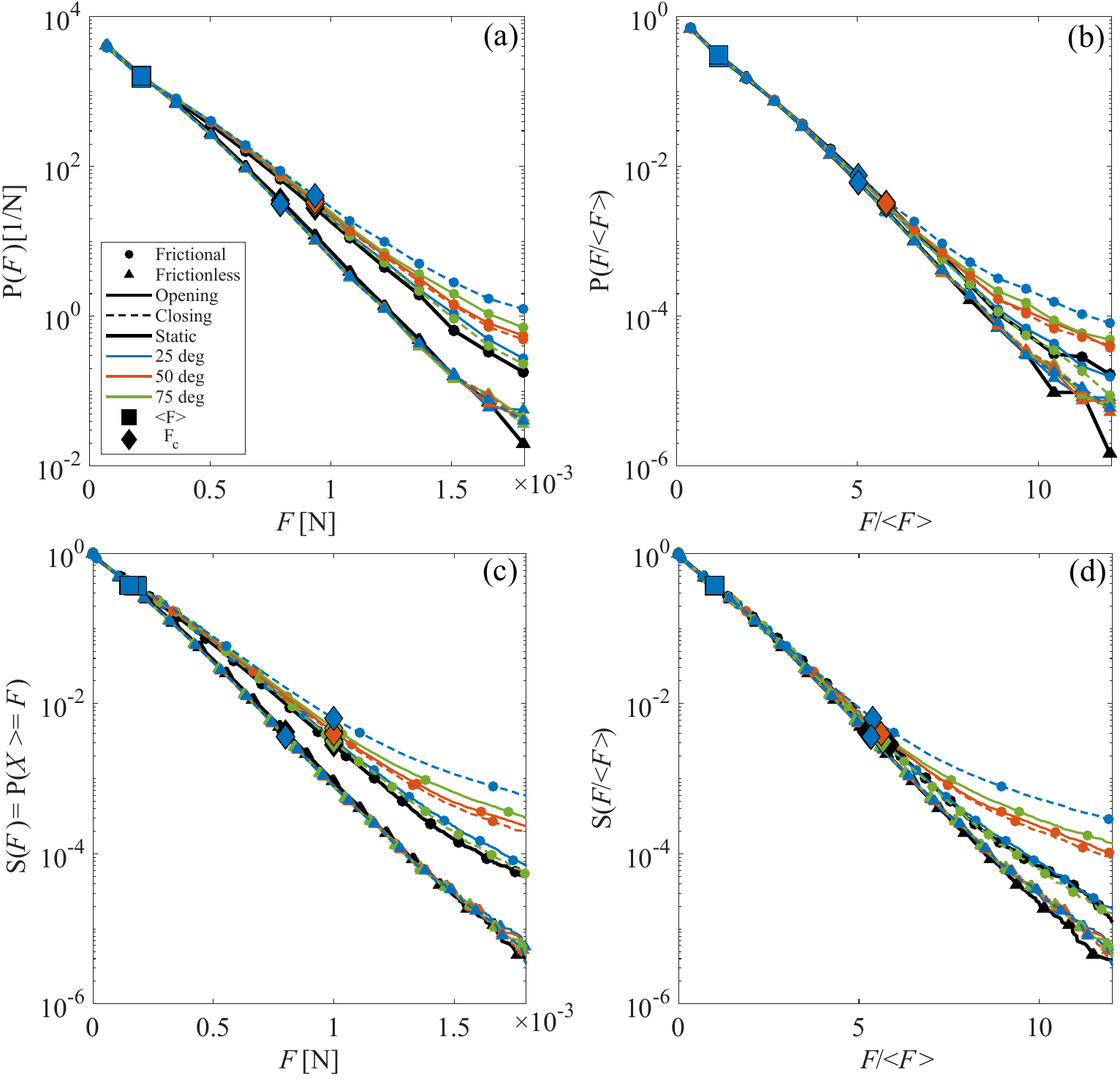}
    \caption{\RR{Probability distributions and complementary cumulative distributions of the contact forces in the static configuration (black) and during the opening (solid lines) and closing (dashed lines) strokes for the base case in the frictional medium (circle markers) and the frictionless medium (triangle markers) at the angles $\theta=25^\circ \text{ (blue)}, \, 50^\circ \text{ (orange)}, \, 75^\circ \text{ (green)}$; (a) $P(F)$; (b) $P(F/\langle F \rangle)$; (c) $S(F)$; (d) $S(F/\langle F \rangle)$. The square markers show $F/\langle F \rangle = 1.0$ and the diamond markers show $F = F_c$ for their corresponding medium. }
    }
    \label{fig:fstat}
\end{figure}

\RR{
To understand the significance and robustness of the chosen force threshold $F_c$ for the $N_c$ calculation, we examine the contact-force statistics by comparing the contact force distribution~\citep{radjai1996force,mueth1998force,silbert2002statistics, snoeijer2004force, majmudar2005contact} in a static reference bed and at representative states during opening and closing. 
More specifically, we study the base case in the frictional medium and the frictionless medium, at the angles $\theta=25^\circ, \, 50^\circ, \, 75^\circ$.
At each example $\theta$, we also compare the opening and closing states, as 
the differences reflect history-dependence of the contact network. 
The resulting probability distributions, $P(F)$, and the complementary cumulative distributions, $S(F)\equiv P(X\geq F)$, of the contact forces are shown in Fig.~\ref{fig:fstat}.} 

\RR{Figure~\ref{fig:fstat}a shows that in the frictional medium, the tail of the distributions at high forces is different between the static bed and the swimming states, indicating swimmer-induced change in the contact force network. 
However, in the frictionless medium, the high-force tails are almost the same between the static and the swimming configurations, indicating little statistically significant change and therefore no hysteresis in the force network.}
\RR{In both media and at all states, the mean force, $\langle F \rangle$, is almost the same, as shown by the square symbols that overlap at $F\approx0.25$\,mN, which is primarily set by the gravitational load. 
To identify a threshold for demarcating the strong forces induced by the swimmer and induced by gravity, we seek a higher threshold than $\langle F \rangle$. In Fig.~\ref{fig:fstat}c, we see that the location at which the distributions differ for the frictional case is $F\approx1$\,mN, which is essentially the top 0.1\% of the contact forces, i.e., our thresholding criterion.
As a result, $N_c$ is a metric that focuses on the far tail of the force distribution, which quantifies the change in the force network due to swimming. As the frictionless base case shows no meaningful changes in the force statistics, $\Delta N_c\approx0$, and no net swimmer displacement occurs. }

\begin{figure}[t]
    \centering
    \includegraphics[width=0.8\linewidth]{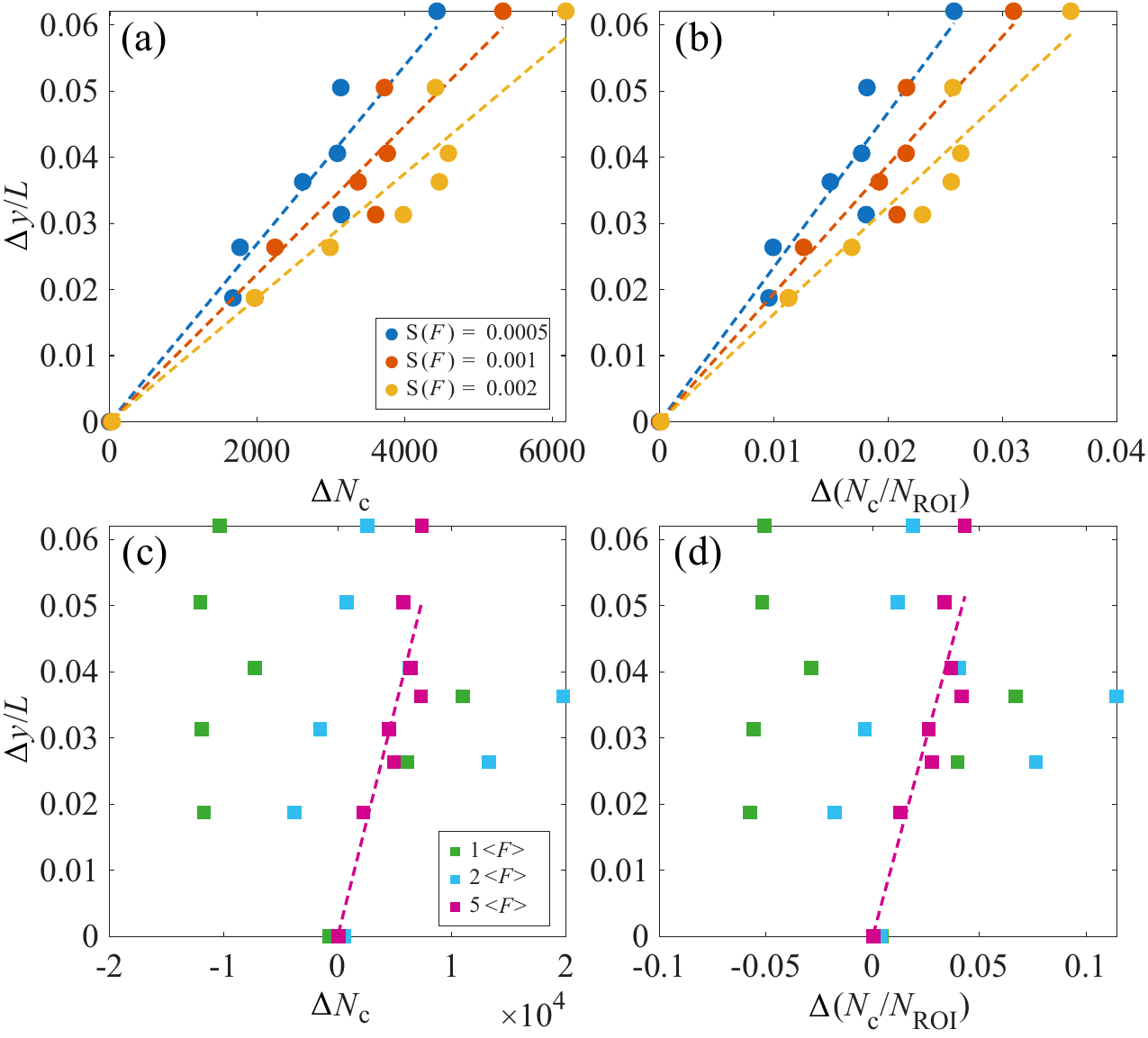}
    \caption{\RR{Relation between the net displacement $\Delta y/L$ and (a, c) $\Delta N_c$ and (b, d) $\Delta(N_c/N_{\mathrm{ROI}})$ for different force thresholds. The complementary cumulative distribution function is used for force thresholding in (a, b), and instantaneous mean force is used in (c, d). The dashed lines show the linear fits to the data. There was no successful linear fit for the instantaneous mean force thresholds $1\langle F \rangle$ and $2\langle F \rangle$.}
    }
    \label{fig:fcrobust}
\end{figure}

\RR{Many prior studies of jamming use $\langle F \rangle$ for thresholding (often in homogeneously loaded granular media)~\citep{snoeijer2004force, majmudar2005contact, radjai1996force, silbert2002statistics}. We show the same distributions but with the contact force normalized by the mean force, $\langle F \rangle$, in Fig.~\ref{fig:fstat}b, which collapses the frictional and frictionless cases. In this case, a reasonable threshold can be found around $F\approx5\langle F \rangle$, whereas in the range $[\langle F \rangle,5\langle F \rangle]$, no significant difference exists. Figure~\ref{fig:fstat}d further shows the corresponding complementary cumulative distributions, indicating that the $5\langle F \rangle$ threshold is essentially equivalent to the top 0.1\% threshold, which is reasonable given that these distributions share a similar exponential form, despite differences at the tails.}

\RR{To further test the robustness of the force threshold $F_c$, 
we show the relation between the net displacement and $\Delta N_c$ for different force thresholds. We use both the instantaneous mean force $F_c=k \langle F \rangle$, where $k=1.0, 2.0, 5.0$ and the inverse of the complementary cumulative distribution function $F_c=S^{-1}(q)$, where $q=0.0005, 0.001, 0.002, $ in Fig.~\ref{fig:fcrobust}a,c. The linear relation between $\Delta N_c$ and $\Delta y/L$ is preserved for large thresholds, including $k=5.0$ and $q=0.0005, 0.001, 0.002$. 
However, no clear relation exists for $k=1.0$ and 2.0. This is because the relatively low thresholds include contact forces that are not strongly related to the swimmer and the associated stagnant zones. 
}

\RR{
We also note that 
using the ratio of the strong contacts to the total contacts in the region of interest, $N_c/N_{\mathrm{ROI}}$, as an intensive metric for the strong contacts still follows the same trend as $N_c$, as shown in Fig.~\ref{fig:fcrobust}b,d, and one can still use $N_c/N_{\mathrm{ROI}}$ as the metric for the strong contacts for other studies.}

\end{appen}

\begin{bmhead}[Acknowledgments.]
The authors would like to acknowledge funding from the National Science Foundation grant CBET-2526568. This research was supported in part through computational resources and services provided by Advanced Research Computing at the University of Michigan, Ann Arbor.
The authors also thank Dr. Thorsten Pöschel for helpful discussions.
\end{bmhead}

\begin{bmhead}[Declaration of interests.]
The authors report no conflicts of interest.
\end{bmhead}

\bibliographystyle{jfm}
\bibliography{ManBib_abbreviated}

\begin{thebibliography}{84}
\expandafter\ifx\csname natexlab\endcsname\relax\def\natexlab#1{#1}\fi
\def\au#1{#1} \def\ed#1{#1} \def\yr#1{#1}\def\at#1{#1}\def\jt#1{\textit{#1}} \def\bt#1{#1}\def\bvol#1{\textbf{#1}} \def\vol#1{#1} \def\pg#1{#1} \def\publ#1{#1}\def\arxiv#1{#1}\def\org#1{#1}\def\st#1{\textit{#1}}

\bibitem[Agarwal {\em et~al.\/}(2023)Agarwal, Goldman \& Kamrin]{agarwal2023mechanistic}
{\sc \au{Agarwal, Shashank}, \au{Goldman, Daniel~I} \& \au{Kamrin, Ken}} \yr{2023}  \at{Mechanistic framework for reduced-order models in soft materials: Application to three-dimensional granular intrusion}.  \jt{Proc. Natl. Acad. Sci. USA}  \bvol{120}~(4),  \pg{e2214017120}.

\bibitem[Agarwal {\em et~al.\/}(2021{\natexlab{{\em a\/}}})Agarwal, Karsai, Goldman \& Kamrin]{agarwal2021efficacy}
{\sc \au{Agarwal, Shashank}, \au{Karsai, Andras}, \au{Goldman, Daniel~I} \& \au{Kamrin, Ken}} \yr{2021{\natexlab{{\em a\/}}}}  \at{Efficacy of simple continuum models for diverse granular intrusions}.  \jt{Soft Matter}  \bvol{17}~(30),  \pg{7196--7209}.

\bibitem[Agarwal {\em et~al.\/}(2021{\natexlab{{\em b\/}}})Agarwal, Karsai, Goldman \& Kamrin]{agarwal2021surprising}
{\sc \au{Agarwal, Shashank}, \au{Karsai, Andras}, \au{Goldman, Daniel~I} \& \au{Kamrin, Ken}} \yr{2021{\natexlab{{\em b\/}}}}  \at{Surprising simplicity in the modeling of dynamic granular intrusion}.  \jt{Sci. Adv.}  \bvol{7}~(17),  \pg{eabe0631}.

\bibitem[Agarwal {\em et~al.\/}(2019)Agarwal, Senatore, Zhang, Kingsbury, Iagnemma, Goldman \& Kamrin]{agarwal2019modeling}
{\sc \au{Agarwal, Shashank}, \au{Senatore, Carmine}, \au{Zhang, Tingnan}, \au{Kingsbury, Mark}, \au{Iagnemma, Karl}, \au{Goldman, Daniel~I} \& \au{Kamrin, Ken}} \yr{2019}  \at{Modeling of the interaction of rigid wheels with dry granular media}.  \jt{J. Terramech.}  \bvol{85},  \pg{1--14}.

\bibitem[Aguilar \& Goldman(2016)]{aguilar2016robophysical}
{\sc \au{Aguilar, Jeffrey} \& \au{Goldman, Daniel~I}} \yr{2016}  \at{Robophysical study of jumping dynamics on granular media}.  \jt{Nat. Phys.}  \bvol{12}~(3),  \pg{278--283}.

\bibitem[Aguilar {\em et~al.\/}(2016)Aguilar, Zhang, Qian, Kingsbury, McInroe, Mazouchova, Li, Maladen, Gong, Travers, Hatton, Choset, Umbanhowar \& Goldman]{aguilar2016review}
{\sc \au{Aguilar, Jeffrey}, \au{Zhang, Tingnan}, \au{Qian, Feifei}, \au{Kingsbury, Mark}, \au{McInroe, Benjamin}, \au{Mazouchova, Nicole}, \au{Li, Chen}, \au{Maladen, Ryan}, \au{Gong, Chaohui}, \au{Travers, Matt}, \au{Hatton, Ross~L}, \au{Choset, Howie}, \au{Umbanhowar, Paul~B} \& \au{Goldman, Daniel~I}} \yr{2016}  \at{A review on locomotion robophysics: the study of movement at the intersection of robotics, soft matter and dynamical systems}.  \jt{Rep. Prog. Phys.}  \bvol{79}~(11),  \pg{110001}.

\bibitem[Albert {\em et~al.\/}(1999)Albert, Pfeifer, Barab{\'a}si \& Schiffer]{albert1999slow}
{\sc \au{Albert, Reka}, \au{Pfeifer, Mark~A}, \au{Barab{\'a}si, Albert-L{\'a}szl{\'o}} \& \au{Schiffer, Peter}} \yr{1999}  \at{Slow drag in a granular medium}.  \jt{Phys. Rev. Lett.}  \bvol{82}~(1),  \pg{205--208}.

\bibitem[Batchelor(2000)]{batchelor2000introduction}
{\sc \au{Batchelor, George~Keith}} \yr{2000} {\em An introduction to fluid dynamics\/}.  \publ{Cambridge university press}.

\bibitem[Becker {\em et~al.\/}(2003)Becker, Koehler \& Stone]{becker2003self}
{\sc \au{Becker, Leif~E}, \au{Koehler, Stephan~A} \& \au{Stone, Howard~A}} \yr{2003}  \at{On self-propulsion of micro-machines at low {Reynolds} number: {Purcell's} three-link swimmer}.  \jt{J. Fluid Mech.}  \bvol{490},  \pg{15--35}.

\bibitem[Behringer \& Chakraborty(2018)]{behringer2018physics}
{\sc \au{Behringer, Robert~P} \& \au{Chakraborty, Bulbul}} \yr{2018}  \at{The physics of jamming for granular materials: a review}.  \jt{Rep. Prog. Phys.}  \bvol{82}~(1),  \pg{012601}.

\bibitem[Bi {\em et~al.\/}(2011)Bi, Zhang, Chakraborty \& Behringer]{bi2011jamming}
{\sc \au{Bi, Dapeng}, \au{Zhang, Jie}, \au{Chakraborty, Bulbul} \& \au{Behringer, Robert~P}} \yr{2011}  \at{Jamming by shear}.  \jt{Nature}  \bvol{480}~(7377),  \pg{355--358}.

\bibitem[Brennen \& Winet(1977)]{brennen1977fluid}
{\sc \au{Brennen, Christopher} \& \au{Winet, Howard}} \yr{1977}  \at{Fluid mechanics of propulsion by cilia and flagella}.  \jt{Annu. Rev. Fluid Mech.}  \bvol{9},  \pg{339--398}.

\bibitem[Brilliantov {\em et~al.\/}(1996)Brilliantov, Spahn, Hertzsch \& P{\"o}schel]{brilliantov1996model}
{\sc \au{Brilliantov, Nikolai~V}, \au{Spahn, Frank}, \au{Hertzsch, Jan-Martin} \& \au{P{\"o}schel, Thorsten}} \yr{1996}  \at{Model for collisions in granular gases}.  \jt{Phys. Rev. E}  \bvol{53}~(5),  \pg{5382}.

\bibitem[Brzinski~III {\em et~al.\/}(2013)Brzinski~III, Mayor \& Durian]{brzinski2013depth}
{\sc \au{Brzinski~III, Theodore~A}, \au{Mayor, Pia} \& \au{Durian, Douglas~J}} \yr{2013}  \at{Depth-dependent resistance of granular media to vertical penetration}.  \jt{Phys. Rev. Lett.}  \bvol{111}~(16),  \pg{168002}.

\bibitem[Chopra {\em et~al.\/}(2020)Chopra, Tolley \& Gravish]{chopra2020granular}
{\sc \au{Chopra, Shivam}, \au{Tolley, Michael~T} \& \au{Gravish, Nick}} \yr{2020}  \at{Granular jamming feet enable improved foot-ground interactions for robot mobility on deformable ground}.  \jt{IEEE Robot. Autom. Lett.}  \bvol{5}~(3),  \pg{3975--3981}.

\bibitem[Clark {\em et~al.\/}(2012)Clark, Kondic \& Behringer]{clark2012particle}
{\sc \au{Clark, Abram~H}, \au{Kondic, Lou} \& \au{Behringer, Robert~P}} \yr{2012}  \at{Particle scale dynamics in granular impact}.  \jt{Phys. Rev. Lett.}  \bvol{109}~(23),  \pg{238302}.

\bibitem[Darbois~Texier {\em et~al.\/}(2021)Darbois~Texier, Ibarra \& Melo]{darbois2021propulsion}
{\sc \au{Darbois~Texier, Baptiste}, \au{Ibarra, Alejandro} \& \au{Melo, Francisco}} \yr{2021}  \at{Propulsion by reciprocal motion into granular media}.  \jt{Phys. Rev. Fluids}  \bvol{6}~(3),  \pg{034604}.

\bibitem[Ding {\em et~al.\/}(2011)Ding, Gravish \& Goldman]{ding2011drag}
{\sc \au{Ding, Yang}, \au{Gravish, Nick} \& \au{Goldman, Daniel~I}} \yr{2011}  \at{Drag induced lift in granular media}.  \jt{Phys. Rev. Lett.}  \bvol{106}~(2),  \pg{028001}.

\bibitem[Ding {\em et~al.\/}(2012)Ding, Sharpe, Masse \& Goldman]{ding2012mechanics}
{\sc \au{Ding, Yang}, \au{Sharpe, Sarah~S}, \au{Masse, Andrew} \& \au{Goldman, Daniel~I}} \yr{2012}  \at{Mechanics of undulatory swimming in a frictional fluid}.  \jt{PLoS Comput. Biol.}  \bvol{8}~(12),  \pg{e1002810}.

\bibitem[Dorgan(2015)]{dorgan2015biomechanics}
{\sc \au{Dorgan, Kelly~M}} \yr{2015}  \at{The biomechanics of burrowing and boring}.  \jt{J. Exp. Biol.}  \bvol{218}~(2),  \pg{176--183}.

\bibitem[Dorgan \& Daltorio(2023)]{dorgan2023fundamentals}
{\sc \au{Dorgan, Kelly~M} \& \au{Daltorio, Kathryn~A}} \yr{2023}  \at{Fundamentals of burrowing in soft animals and robots}.  \jt{Front. Robot. AI}  \bvol{10},  \pg{1057876}.

\bibitem[Feng {\em et~al.\/}(2019)Feng, Blumenfeld \& Liu]{feng2019support}
{\sc \au{Feng, Yajie}, \au{Blumenfeld, Raphael} \& \au{Liu, Caishan}} \yr{2019}  \at{Support of modified {Archimedes'} law theory in granular media}.  \jt{Soft Matter}  \bvol{15}~(14),  \pg{3008--3017}.

\bibitem[de~Gennes(1999)]{de1999granular}
{\sc \au{de~Gennes, Pierre-Gilles}} \yr{1999}  \at{Granular matter: a tentative view}.  \jt{Rev. Mod. Phys.}  \bvol{71}~(2),  \pg{S374}.

\bibitem[Gonzalez-Rodriguez \& Lauga(2009)]{gonzalez2009reciprocal}
{\sc \au{Gonzalez-Rodriguez, David} \& \au{Lauga, Eric}} \yr{2009}  \at{Reciprocal locomotion of dense swimmers in {Stokes} flow}.  \jt{J. Phys. Condens. Matter}  \bvol{21}~(20),  \pg{204103}.

\bibitem[Harrington {\em et~al.\/}(2020)Harrington, Xiao \& Durian]{harrington2020stagnant}
{\sc \au{Harrington, Matt}, \au{Xiao, Hongyi} \& \au{Durian, Douglas~J}} \yr{2020}  \at{Stagnant zone formation in a {2D} bed of circular and elongated grains under penetration}.  \jt{Granul. Matter}  \bvol{22},  \pg{1--9}.

\bibitem[Hertz(1881)]{hertz1882z}
{\sc \au{Hertz, Heinrich}} \yr{1881}  \at{{\"U}ber die {Ber{\"u}hrung} fester elastischer {K{\"o}rper}}.  \jt{J. fur die reine u. angew. Mathem.}  \bvol{92},  \pg{156}.

\bibitem[Hilton \& Tordesillas(2013)]{hilton2013drag}
{\sc \au{Hilton, James~E} \& \au{Tordesillas, Antoinette}} \yr{2013}  \at{Drag force on a spherical intruder in a granular bed at low {F}roude number}.  \jt{Phys. Rev. E}  \bvol{88}~(6),  \pg{062203}.

\bibitem[Holm \& Edney(1973)]{holm1973daily}
{\sc \au{Holm, Erik} \& \au{Edney, EB}} \yr{1973}  \at{Daily activity of {Namib} desert arthropods in relation to climate}.  \jt{Ecology}  \bvol{54}~(1),  \pg{45--56}.

\bibitem[Hosoi \& Goldman(2015)]{hosoi2015beneath}
{\sc \au{Hosoi, AE} \& \au{Goldman, Daniel~I}} \yr{2015}  \at{Beneath our feet: strategies for locomotion in granular media}.  \jt{Annu. Rev. Fluid Mech.}  \bvol{47}~(1),  \pg{431--453}.

\bibitem[Hubert {\em et~al.\/}(2021)Hubert, Trosman, Collard, Sukhov, Harting, Vandewalle \& Smith]{hubert2021scallop}
{\sc \au{Hubert, Maxime}, \au{Trosman, Oleg}, \au{Collard, Ylona}, \au{Sukhov, Alexander}, \au{Harting, Jens}, \au{Vandewalle, Nicolas} \& \au{Smith, A-S}} \yr{2021}  \at{Scallop theorem and swimming at the mesoscale}.  \jt{Phys. Rev. Lett.}  \bvol{126}~(22),  \pg{224501}.

\bibitem[Isaka {\em et~al.\/}(2019)Isaka, Tsumura, Watanabe, Toyama, Sugesawa, Yamada, Yoshida \& Nakamura]{isaka2019development}
{\sc \au{Isaka, Keita}, \au{Tsumura, Kazuki}, \au{Watanabe, Tomoki}, \au{Toyama, Wataru}, \au{Sugesawa, Makoto}, \au{Yamada, Yasuyuki}, \au{Yoshida, Hiroshi} \& \au{Nakamura, Taro}} \yr{2019}  \at{Development of underwater drilling robot based on earthworm locomotion}.  \jt{IEEE Access}  \bvol{7},  \pg{103127--103141}.

\bibitem[Jaeger {\em et~al.\/}(1996)Jaeger, Nagel \& Behringer]{jaeger1996granular}
{\sc \au{Jaeger, Heinrich~M}, \au{Nagel, Sidney~R} \& \au{Behringer, Robert~P}} \yr{1996}  \at{Granular solids, liquids, and gases}.  \jt{Rev. Mod. Phys.}  \bvol{68}~(4),  \pg{1259}.

\bibitem[Jop {\em et~al.\/}(2006)Jop, Forterre \& Pouliquen]{jop2006constitutive}
{\sc \au{Jop, Pierre}, \au{Forterre, Yo{\"e}l} \& \au{Pouliquen, Olivier}} \yr{2006}  \at{A constitutive law for dense granular flows}.  \jt{Nature}  \bvol{441}~(7094),  \pg{727--730}.

\bibitem[Kamrin {\em et~al.\/}(2024)Kamrin, Hill, Goldman \& Andrade]{kamrin2024advances}
{\sc \au{Kamrin, Ken}, \au{Hill, Kimberly~M}, \au{Goldman, Daniel~I} \& \au{Andrade, Jose~E}} \yr{2024}  \at{Advances in modeling dense granular media}.  \jt{Annu. Rev. Fluid Mech.}  \bvol{56}~(1),  \pg{215--240}.

\bibitem[Kamrin \& Koval(2012)]{kamrin2012nonlocal}
{\sc \au{Kamrin, Ken} \& \au{Koval, Georg}} \yr{2012}  \at{Nonlocal constitutive relation for steady granular flow}.  \jt{Phys. Rev. Lett.}  \bvol{108}~(17),  \pg{178301}.

\bibitem[Kang {\em et~al.\/}(2018)Kang, Feng, Liu \& Blumenfeld]{kang2018archimedes}
{\sc \au{Kang, Wenting}, \au{Feng, Yajie}, \au{Liu, Caishan} \& \au{Blumenfeld, Raphael}} \yr{2018}  \at{{Archimedes’} law explains penetration of solids into granular media}.  \jt{Nat. Commun.}  \bvol{9}~(1),  \pg{1101}.

\bibitem[Katsuragi \& Durian(2007)]{katsuragi2007unified}
{\sc \au{Katsuragi, Hiroaki} \& \au{Durian, Douglas~J}} \yr{2007}  \at{Unified force law for granular impact cratering}.  \jt{Nat. Phys.}  \bvol{3}~(6),  \pg{420--423}.

\bibitem[Koiller {\em et~al.\/}(1996)Koiller, Ehlers \& Montgomery]{koiller1996problems}
{\sc \au{Koiller, Jair}, \au{Ehlers, Kurt} \& \au{Montgomery, Richard}} \yr{1996}  \at{Problems and progress in microswimming}.  \jt{J. Nonlinear Sci.}  \bvol{6}~(6),  \pg{507--541}.

\bibitem[Kozlowski {\em et~al.\/}(2019)Kozlowski, Carlevaro, Daniels, Kondic, Pugnaloni, Socolar, Zheng \& Behringer]{kozlowski2019dynamics}
{\sc \au{Kozlowski, Ryan}, \au{Carlevaro, C~Manuel}, \au{Daniels, Karen~E}, \au{Kondic, Lou}, \au{Pugnaloni, Luis~A}, \au{Socolar, Joshua~ES}, \au{Zheng, Hu} \& \au{Behringer, Robert~P}} \yr{2019}  \at{Dynamics of a grain-scale intruder in a two-dimensional granular medium with and without basal friction}.  \jt{Phys. Rev. E}  \bvol{100}~(3),  \pg{032905}.

\bibitem[Lauga(2011)]{lauga2011life}
{\sc \au{Lauga, Eric}} \yr{2011}  \at{Life around the scallop theorem}.  \jt{Soft Matter}  \bvol{7}~(7),  \pg{3060--3065}.

\bibitem[Lauga \& Bartolo(2008)]{lauga2008no}
{\sc \au{Lauga, Eric} \& \au{Bartolo, Denis}} \yr{2008}  \at{No many-scallop theorem: Collective locomotion of reciprocal swimmers}.  \jt{Phys. Rev. E}  \bvol{78}~(3),  \pg{030901}.

\bibitem[Li {\em et~al.\/}(2013)Li, Zhang \& Goldman]{li2013terradynamics}
{\sc \au{Li, Chen}, \au{Zhang, Tingnan} \& \au{Goldman, Daniel~I}} \yr{2013}  \at{A terradynamics of legged locomotion on granular media}.  \jt{Science}  \bvol{339}~(6126),  \pg{1408--1412}.

\bibitem[Li {\em et~al.\/}(2021)Li, Huang, Tang, Marvi, Tao \& Aukes]{li2021compliant}
{\sc \au{Li, Dongting}, \au{Huang, Sichuan}, \au{Tang, Yong}, \au{Marvi, Hamidreza}, \au{Tao, Junliang} \& \au{Aukes, Daniel~M}} \yr{2021}  \at{Compliant fins for locomotion in granular media}.  \jt{IEEE Robot. Autom. Lett.}  \bvol{6}~(3),  \pg{5984--5991}.

\bibitem[Li {\em et~al.\/}(2024)Li, Zhao, He, Qi, Kang \& Ma]{li2024enhancing}
{\sc \au{Li, Longchuan}, \au{Zhao, Chaoyue}, \au{He, Shuqian}, \au{Qi, Qiukai}, \au{Kang, Shuai} \& \au{Ma, Shugen}} \yr{2024}  \at{Enhancing undulation of soft robots in granular media: A numerical and experimental study on the effect of anisotropic scales}.  \jt{Biomim. Intell. Robot.}  \bvol{4}~(2),  \pg{100158}.

\bibitem[Lighthill(1976)]{lighthill1976flagellar}
{\sc \au{Lighthill, James}} \yr{1976}  \at{Flagellar hydrodynamics}.  \jt{SIAM Rev.}  \bvol{18}~(2),  \pg{161--230}.

\bibitem[Majmudar \& Behringer(2005)]{majmudar2005contact}
{\sc \au{Majmudar, Trushant~S} \& \au{Behringer, Robert~P}} \yr{2005}  \at{Contact force measurements and stress-induced anisotropy in granular materials}.  \jt{Nature}  \bvol{435}~(7045),  \pg{1079--1082}.

\bibitem[Maladen {\em et~al.\/}(2009)Maladen, Ding, Li \& Goldman]{maladen2009undulatory}
{\sc \au{Maladen, Ryan~D}, \au{Ding, Yang}, \au{Li, Chen} \& \au{Goldman, Daniel~I}} \yr{2009}  \at{Undulatory swimming in sand: subsurface locomotion of the sandfish lizard}.  \jt{Science}  \bvol{325}~(5938),  \pg{314--318}.

\bibitem[Maladen {\em et~al.\/}(2011)Maladen, Ding, Umbanhowar, Kamor \& Goldman]{maladen2011mechanical}
{\sc \au{Maladen, Ryan~D}, \au{Ding, Yang}, \au{Umbanhowar, Paul~B}, \au{Kamor, Adam} \& \au{Goldman, Daniel~I}} \yr{2011}  \at{Mechanical models of sandfish locomotion reveal principles of high performance subsurface sand-swimming}.  \jt{J. R. Soc. Interface}  \bvol{8}~(62),  \pg{1332--1345}.

\bibitem[MiDi(2004)]{gdr2004dense}
{\sc \au{MiDi, GDR}} \yr{2004}  \at{On dense granular flows}.  \jt{Eur. Phys. J. E}  \bvol{14}~(4),  \pg{341--365}.

\bibitem[Mindlin(1949)]{mindlin1949compliance}
{\sc \au{Mindlin, RD}} \yr{1949}  \at{Compliance of elastic bodies in contact}.  \jt{J. Appl. Mech.}  \bvol{16}~(3),  \pg{259--268}.

\bibitem[Mueth {\em et~al.\/}(1998)Mueth, Jaeger \& Nagel]{mueth1998force}
{\sc \au{Mueth, Daniel~M}, \au{Jaeger, Heinrich~M} \& \au{Nagel, Sidney~R}} \yr{1998}  \at{Force distribution in a granular medium}.  \jt{Phys. Rev. E}  \bvol{57}~(3),  \pg{3164--3169}.

\bibitem[M{\"u}ller \& P{\"o}schel(2011)]{muller2011collision}
{\sc \au{M{\"u}ller, Patric} \& \au{P{\"o}schel, Thorsten}} \yr{2011}  \at{Collision of viscoelastic spheres: Compact expressions for the coefficient of normal restitution}.  \jt{Phys. Rev. E}  \bvol{84}~(2),  \pg{021302}.

\bibitem[Otsuki \& Hayakawa(2021)]{otsuki2021shear}
{\sc \au{Otsuki, Michio} \& \au{Hayakawa, Hisao}} \yr{2021}  \at{Shear modulus and reversible particle trajectories of frictional granular materials under oscillatory shear}.  \jt{Eur. Phys. J. E}  \bvol{44}~(5),  \pg{70}.

\bibitem[Peng {\em et~al.\/}(2016)Peng, Pak \& Elfring]{peng2016characteristics}
{\sc \au{Peng, Zhiwei}, \au{Pak, On~Shun} \& \au{Elfring, Gwynn~J}} \yr{2016}  \at{Characteristics of undulatory locomotion in granular media}.  \jt{Phys. Fluids}  \bvol{28}~(3),  \pg{031901}.

\bibitem[Pravin {\em et~al.\/}(2021)Pravin, Chang, Han, London, Goldman, Jaeger \& Hsieh]{pravin2021effect}
{\sc \au{Pravin, Swapnil}, \au{Chang, Brian}, \au{Han, Endao}, \au{London, Lionel}, \au{Goldman, Daniel~I}, \au{Jaeger, Heinrich~M} \& \au{Hsieh, S~Tonia}} \yr{2021}  \at{Effect of two parallel intruders on total work during granular penetrations}.  \jt{Phys. Rev. E}  \bvol{104}~(2),  \pg{024902}.

\bibitem[Purcell(1977)]{purcell2014life}
{\sc \au{Purcell, Edward~M}} \yr{1977}  \at{Life at low {Reynolds} number}.  \jt{Am. J. Phys.}  \bvol{45}~(1),  \pg{3--11}.

\bibitem[Radja{\"i} {\em et~al.\/}(1996)Radja{\"i}, Jean, Moreau \& Roux]{radjai1996force}
{\sc \au{Radja{\"i}, Farhang}, \au{Jean, Michel}, \au{Moreau, Jean-Jacques} \& \au{Roux, Stephane}} \yr{1996}  \at{Force distributions in dense two-dimensional granular systems}.  \jt{Phys. Rev. Lett.}  \bvol{77}~(2),  \pg{274--277}.

\bibitem[van Rees {\em et~al.\/}(2015)van Rees, Novati \& Koumoutsakos]{van2015self}
{\sc \au{van Rees, Wim~M}, \au{Novati, Guido} \& \au{Koumoutsakos, Petros}} \yr{2015}  \at{Self-propulsion of a counter-rotating cylinder pair in a viscous fluid}.  \jt{Phys. Fluids}  \bvol{27}~(6),  \pg{063102}.

\bibitem[Robertson {\em et~al.\/}(2019)Robertson, Efremov \& Paik]{robertson2019roboscallop}
{\sc \au{Robertson, Matthew~A}, \au{Efremov, Filip} \& \au{Paik, Jamie}} \yr{2019}  \at{Roboscallop: A bivalve inspired swimming robot}.  \jt{IEEE Robot. Autom. Lett.}  \bvol{4}~(2),  \pg{2078--2085}.

\bibitem[Roy {\em et~al.\/}(2024{\natexlab{{\em a\/}}})Roy, Xiao, Angelidakis \& P{\"o}schel]{roy2024combined}
{\sc \au{Roy, Sudeshna}, \au{Xiao, Hongyi}, \au{Angelidakis, Vasileios} \& \au{P{\"o}schel, Thorsten}} \yr{2024{\natexlab{{\em a\/}}}}  \at{Combined thermal and particle shape effects on powder spreading in additive manufacturing via discrete element simulations}.  \jt{Powder Technol.}  \bvol{445},  \pg{120099}.

\bibitem[Roy {\em et~al.\/}(2024{\natexlab{{\em b\/}}})Roy, Xiao, Angelidakis \& P{\"o}schel]{roy2024structural}
{\sc \au{Roy, Sudeshna}, \au{Xiao, Hongyi}, \au{Angelidakis, Vasileios} \& \au{P{\"o}schel, Thorsten}} \yr{2024{\natexlab{{\em b\/}}}}  \at{Structural fluctuations in thin cohesive particle layers in powder-based additive manufacturing}.  \jt{Granul. Matter}  \bvol{26}~(2),  \pg{43}.

\bibitem[Schiebel {\em et~al.\/}(2020)Schiebel, Astley, Rieser, Agarwal, Hubicki, Hubbard, Diaz, Mendelson~III, Kamrin \& Goldman]{schiebel2020mitigating}
{\sc \au{Schiebel, Perrin~E}, \au{Astley, Henry~C}, \au{Rieser, Jennifer~M}, \au{Agarwal, Shashank}, \au{Hubicki, Christian}, \au{Hubbard, Alex~M}, \au{Diaz, Kelimar}, \au{Mendelson~III, Joseph~R}, \au{Kamrin, Ken} \& \au{Goldman, Daniel~I}} \yr{2020}  \at{Mitigating memory effects during undulatory locomotion on hysteretic materials}.  \jt{eLife}  \bvol{9},  \pg{e51412}.

\bibitem[Shapere \& Wilczek(1987)]{shapere1987self}
{\sc \au{Shapere, Alfred} \& \au{Wilczek, Frank}} \yr{1987}  \at{Self-propulsion at low {Reynolds} number}.  \jt{Phys. Rev. Lett.}  \bvol{58}~(20),  \pg{2051--2054}.

\bibitem[Shapere \& Wilczek(1989)]{shapere1989geometry}
{\sc \au{Shapere, Alfred} \& \au{Wilczek, Frank}} \yr{1989}  \at{Geometry of self-propulsion at low {Reynolds} number}.  \jt{J. Fluid Mech.}  \bvol{198},  \pg{557--585}.

\bibitem[Silbert(2010)]{silbert2010jamming}
{\sc \au{Silbert, Leonardo~E}} \yr{2010}  \at{Jamming of frictional spheres and random loose packing}.  \jt{Soft Matter}  \bvol{6}~(13),  \pg{2918--2924}.

\bibitem[Silbert {\em et~al.\/}(2002)Silbert, Grest \& Landry]{silbert2002statistics}
{\sc \au{Silbert, Leonardo~E}, \au{Grest, Gary~S} \& \au{Landry, James~W}} \yr{2002}  \at{Statistics of the contact network in frictional and frictionless granular packings}.  \jt{Phys. Rev. E}  \bvol{66}~(6),  \pg{061303}.

\bibitem[Snoeijer {\em et~al.\/}(2004)Snoeijer, Vlugt, van Hecke \& van Saarloos]{snoeijer2004force}
{\sc \au{Snoeijer, Jacco~H}, \au{Vlugt, Thijs J~H}, \au{van Hecke, Martin} \& \au{van Saarloos, Wim}} \yr{2004}  \at{Force network ensemble: a new approach to static granular matter}.  \jt{Phys. Rev. Lett.}  \bvol{92}~(5),  \pg{054302}.

\bibitem[Strobl {\em et~al.\/}(2016)Strobl, Formella \& P{\"o}schel]{strobl2016exact}
{\sc \au{Strobl, Severin}, \au{Formella, Arno} \& \au{P{\"o}schel, Thorsten}} \yr{2016}  \at{Exact calculation of the overlap volume of spheres and mesh elements}.  \jt{J. Comput. Phys.}  \bvol{311},  \pg{158--172}.

\bibitem[Takehara \& Okumura(2014)]{takehara2014friction}
{\sc \au{Takehara, Yousuke} \& \au{Okumura, Ko}} \yr{2014}  \at{High-velocity drag friction in granular media near the jamming point}.  \jt{Phys. Rev. Lett.}  \bvol{112}~(14),  \pg{148001}.

\bibitem[Thornton {\em et~al.\/}(2023)Thornton, Plath, Ostanin, G{\"o}tz, Bisschop, Hassan, Roeplal, Wang, Pourandi \& Weinhart]{thornton2023recent}
{\sc \au{Thornton, Anthony~R}, \au{Plath, Timo}, \au{Ostanin, Igor}, \au{G{\"o}tz, Holger}, \au{Bisschop, Jan-Willem}, \au{Hassan, Mohamed}, \au{Roeplal, Ra{\"\i}sa}, \au{Wang, Xiuqi}, \au{Pourandi, Sahar} \& \au{Weinhart, Thomas}} \yr{2023}  \at{Recent advances in {MercuryDPM}}.  \jt{Math. Comput. Sci.}  \bvol{17}~(2),  \pg{13}.

\bibitem[Thornton(2015)]{thornton2015granular}
{\sc \au{Thornton, Colin}} \yr{2015}  \at{Granular dynamics, contact mechanics and particle system simulations}.  \jt{Part. Technol. Ser.}  \bvol{24}.

\bibitem[Verlet(1967)]{verlet1967computer}
{\sc \au{Verlet, Loup}} \yr{1967}  \at{Computer ``experiments" on classical fluids. i. thermodynamical properties of {Lennard-Jones} molecules}.  \jt{Phys. Rev.}  \bvol{159}~(1),  \pg{98--103}.

\bibitem[Wang {\em et~al.\/}(2020)Wang, Pang, Jin, Xu, Sun, Li \& Zhang]{wang2020development}
{\sc \au{Wang, Yumo}, \au{Pang, Shunxiang}, \au{Jin, Hu}, \au{Xu, Min}, \au{Sun, Shuaishuai}, \au{Li, Weihua} \& \au{Zhang, Shiwu}} \yr{2020}  \at{Development of a biomimetic scallop robot capable of jet propulsion}.  \jt{Bioinspir. Biomim.}  \bvol{15}~(3),  \pg{036008}.

\bibitem[Wang {\em et~al.\/}(2018)Wang, Sun, Xu, Li \& Zhang]{wang2018design}
{\sc \au{Wang, Yumo}, \au{Sun, Shuaishuai}, \au{Xu, Min}, \au{Li, Weihua} \& \au{Zhang, Shiwu}} \yr{2018} Design of a bionic scallop robot based on jet propulsion.  \bt{In {\em 2018 IEEE International Conference on Real-time Computing and Robotics (RCAR)\/}},  \pg{pp. 563--566}. IEEE.

\bibitem[Weinhart {\em et~al.\/}(2020)Weinhart, Orefice, Post, {van Schrojenstein Lantman}, Denissen, Tunuguntla, Tsang, Cheng, Shaheen, Shi, Rapino, Grannonio, Losacco, Barbosa, Jing, {Alvarez Naranjo}, Roy, {den Otter} \& Thornton]{WEINHART2020107129}
{\sc \au{Weinhart, Thomas}, \au{Orefice, Luca}, \au{Post, Mitchel}, \au{{van Schrojenstein Lantman}, Marnix~P.}, \au{Denissen, Irana~F.C.}, \au{Tunuguntla, Deepak~R.}, \au{Tsang, J.M.F.}, \au{Cheng, Hongyang}, \au{Shaheen, Mohamad~Yousef}, \au{Shi, Hao}, \au{Rapino, Paolo}, \au{Grannonio, Elena}, \au{Losacco, Nunzio}, \au{Barbosa, Joao}, \au{Jing, Lu}, \au{{Alvarez Naranjo}, Juan~E.}, \au{Roy, Sudeshna}, \au{{den Otter}, Wouter~K.} \& \au{Thornton, Anthony~R.}} \yr{2020}  \at{Fast, flexible particle simulations — an introduction to {MercuryDPM}}.  \jt{Comput. Phys. Commun.}  \bvol{249},  \pg{107129}.

\bibitem[Winter {\em et~al.\/}(2014)Winter, Deits, Dorsch, Slocum \& Hosoi]{winter2014razor}
{\sc \au{Winter, AG}, \au{Deits, RLH}, \au{Dorsch, DS}, \au{Slocum, AH} \& \au{Hosoi, AE}} \yr{2014}  \at{Razor clam to roboclam: burrowing drag reduction mechanisms and their robotic adaptation}.  \jt{Bioinspir. Biomim.}  \bvol{9}~(3),  \pg{036009}.

\bibitem[Xiao {\em et~al.\/}(2025)Xiao, Torres, Sack \& P{\"o}schel]{xiao2024locomotion}
{\sc \au{Xiao, Hongyi}, \au{Torres, Harol}, \au{Sack, Achim} \& \au{P{\"o}schel, Thorsten}} \yr{2025}  \at{Locomotion of a scallop-inspired swimmer in granular matter}.  \jt{Phys. Rev. Appl.}  \bvol{24}~(3),  \pg{034049}.

\bibitem[Y{\i}lmaz {\em et~al.\/}(2026)Y{\i}lmaz, Douglas, Perron \& Kamrin]{Yilmaz2026ElasticRFT}
{\sc \au{Y{\i}lmaz, Lale}, \au{Douglas, Madison}, \au{Perron, J.~Taylor} \& \au{Kamrin, Ken}} \yr{2026}  \at{Elastic resistive force theory: development and applications to flexible intruders}.  \jt{Soft Matter}  \bvol{22}~(14),  \pg{2638--2655}.

\bibitem[Yin {\em et~al.\/}(2025)Yin, Huang, Yu \& Liu]{yin2025extended}
{\sc \au{Yin, Yuefan}, \au{Huang, Shuo}, \au{Yu, Yang} \& \au{Liu, Caishan}} \yr{2025}  \at{Extended application of modified {Archimedes’} law in granular media}.  \jt{Powder Technol.}  \bvol{452},  \pg{120560}.

\bibitem[Zhang {\em et~al.\/}(2010)Zhang, Majmudar, Tordesillas \& Behringer]{zhang2010statistical}
{\sc \au{Zhang, J}, \au{Majmudar, TS}, \au{Tordesillas, A} \& \au{Behringer, RP}} \yr{2010}  \at{Statistical properties of a {2D} granular material subjected to cyclic shear}.  \jt{Granul. Matter}  \bvol{12},  \pg{159--172}.

\bibitem[Zhang {\em et~al.\/}(2023)Zhang, Chen, Martinez \& Fuentes]{zhang2023bioinspired}
{\sc \au{Zhang, Ningning}, \au{Chen, Yuyan}, \au{Martinez, Alejandro} \& \au{Fuentes, Raul}} \yr{2023}  \at{A bioinspired self-burrowing probe in shallow granular materials}.  \jt{J. Geotech. Geoenviron. Eng.}  \bvol{149}~(9),  \pg{04023073}.

\bibitem[Zhang {\em et~al.\/}(2019)Zhang, Cao, Wang, Wang, Zhu \& Zhang]{zhang2019motion}
{\sc \au{Zhang, Yan}, \au{Cao, Jiafeng}, \au{Wang, Qi}, \au{Wang, Pengfei}, \au{Zhu, Yueying} \& \au{Zhang, Junxia}} \yr{2019}  \at{Motion characteristics of the appendages of mole crickets during burrowing}.  \jt{J. Bionic Eng.}  \bvol{16},  \pg{319--327}.

\bibitem[Zhao {\em et~al.\/}(2019)Zhao, Bar{\'e}s, Zheng, Socolar \& Behringer]{zhao2019shear}
{\sc \au{Zhao, Yiqiu}, \au{Bar{\'e}s, Jonathan}, \au{Zheng, Hu}, \au{Socolar, Joshua~ES} \& \au{Behringer, Robert~P}} \yr{2019}  \at{Shear-jammed, fragile, and steady states in homogeneously strained granular materials}.  \jt{Phys. Rev. Lett.}  \bvol{123}~(15),  \pg{158001}.

\bibitem[Zhao {\em et~al.\/}(2022)Zhao, Zhao, Wang, Zheng, Chakraborty \& Socolar]{zhao2022ultrastable}
{\sc \au{Zhao, Yiqiu}, \au{Zhao, Yuchen}, \au{Wang, Dong}, \au{Zheng, Hu}, \au{Chakraborty, Bulbul} \& \au{Socolar, Joshua~ES}} \yr{2022}  \at{Ultrastable shear-jammed granular material}.  \jt{Phys. Rev. X}  \bvol{12}~(3),  \pg{031021}.

\end{thebibliography}


\end{document}